\documentclass[pra,twocolumn,english,aps,showpacs,superscriptaddress,amsmath,amsfonts]{revtex4-1}

\usepackage{graphicx}
\usepackage{color}

\usepackage{bm, bbm}      
\usepackage{subfigure}

\usepackage{amssymb}

\usepackage[english]{babel}

\usepackage{textcomp}

\usepackage{array}
\newcolumntype{L}[1]{>{\raggedright\let\newline\\\arraybackslash\hspace{0pt}}m{#1}}
\newcolumntype{C}[1]{>{\centering\let\newline\\\arraybackslash\hspace{0pt}}m{#1}}
\newcolumntype{R}[1]{>{\raggedleft\let\newline\\\arraybackslash\hspace{0pt}}m{#1}}

\begin{document}

\title{Spin-orbital order in undoped manganite LaMnO$_3$ at finite temperature}

\author     {Mateusz Snamina}
\affiliation{Kazimierz Gumi\'nski Department of Theoretical Chemistry, Faculty of Chemistry,
             Jagiellonian University, prof. R. Ingardena 3, PL-30060 Krak\'ow, Poland }

\author{     Andrzej M. Ole\'s}
\affiliation{Marian Smoluchowski Institute of Physics, Jagiellonian
             University, prof. S. \L{}ojasiewicza 11, PL-30348 Krak\'ow, Poland }
\affiliation{Max Planck Institute for Solid State Research,
             Heisenbergstrasse 1, D-70569 Stuttgart, Germany}

\date{1 November 2016}

\begin{abstract}
We investigate the evolution of spin and orbital order in undoped
LaMnO$_3$ under increasing temperature with a model including both
superexchange and Jahn-Teller interactions. We used several cluster
mean field calculation schemes and find coexisting $A$-type
antiferromagnetic ($A$-AF) and $C$-type alternating orbital order
at low temperature. The value of the Jahn-Teller coupling between
strongly correlated $e_g$ orbitals is estimated from the orbital
transition temperature at $T_{\rm OO}\simeq 780$ K. By a careful
analysis of on-site and on-bond correlations we demonstrate that
spin-orbital entanglement is rather weak.
We have verified that the magnetic transition temperature is
influenced by entangled spin-orbital operators as well as by
entangled orbital operators on the bonds but the errors introduced
by decoupling such operators partly compensate each other.
Altogether, these results justify why the commonly used disentangled
spin-orbital model is so successful in describing the magnetic
properties and the temperature dependence of the optical spectral
weights for LaMnO$_3$.\\
\textit{Published in: Physical Review B \textbf{94}, 214426 (2016).}
\end{abstract}


\maketitle

\section{Introduction}

Recent extensive work on transition metal oxides has demonstrated
a strong interrelationship between spin order and orbital order,
resulting in phase diagrams of great complexity for undoped compounds
with active orbital degrees of freedom \cite{Tok00}. A good example
is LaMnO$_3$, the parent compound of colossal magnetoresistance,
where thorough experimental work has produced exceptionally detailed
information on the phase diagrams of the $R$MnO$_3$ manganites (where
$R$=Lu, Yb,$\dots$, La), yet the interplay between spin and orbital
degrees of freedom is puzzling and could not be resolved by the theory
so far \cite{Goode}. This difficulty follows in general from the
coupling to the lattice via the Jahn-Teller (JT) effect and partly from
spin-orbital quantum fluctuations \cite{Fei97,Kha00,Kha01,Kha04,Kha05}
which contribute to superexchange \cite{Kug82,Ole05,Hor08,Woh11,Brz15}
and are amplified in the presence of spin-orbital entanglement on
superexchange bonds \cite{Ole12,Rex12,You15}. The prominent example are
the perovskite vanadium oxides where the temperature dependence of
optical spectral weights \cite{Miy02} and the phase diagram
\cite{Fuj10} could be understood in the theory only by an explicit
treatment of entanglement \cite{Kha04,Hor08}. It is therefore of
fundamental importance to consider simultaneously quantum spin and
orbital degrees of freedom when investigating the possibility of
spin-orbital order.

Although ordered spin-orbital states occur in many cases including
LaMnO$_3$ \cite{Dag01,Tok06}, disordered quantum phases are very
challenging, such as spin \cite{Bal10,Bal16} or orbital
\cite{Kha00,Fei05} liquids when one of the two degrees of freedom is
frozen. Another disordered state is a quantum spin-orbital liquid
\cite{Nor08,Karlo,Nas12,Sel14,Mil14}, where
spin-orbital order is absent and occupied (by electrons or holes)
spin-orbital states are randomly chosen by electrons. Previous attempts
to find a spin-orbital liquid in the Kugel-Khomskii model \cite{Fei97}
or in LiNiO$_2$ \cite{Ver04} turned out to be unsuccessful, and it was
established that instead:
(i) novel types of exotic spin-orbital order emerge from entanglement
in the Kugel-Khomskii model \cite{Brz12}, and
(ii) spin and orbital interactions are of quite different strengths in
LiNiO$_2$ and the reasons behind the absence of magnetic long range
order are more subtle \cite{Rei05}.
Therefore, the right strategy is to investigate whether ordered states
may occur and to what extent strong spin-orbital coupling (including
possible entanglement) influences the spin and orbital phase transitions.

Here we focus on LaMnO$_3$, a Mott insulator with Mn$^{3+}$ ions in
$t_{2g}^3e_g^1$ configuration with high $S=2$ spin state stabilized by
Hund's exchange. The singly occupied $e_g$ orbitals are the orbital
degrees of freedom which order below an orbital transition at
$T_{\rm OO}\simeq 780$ K, recognized as a strong JT instability when
cooperative distortions of oxygen octahedra occur and induce
alternating orbital (AO) order \cite{Goo55,Fei98,Nan10}.
Antiferromagnetic (AF) order of $A$-type ($A$-AF), i.e., AF phase with
ferromagnetic (FM) $(a,b)$ planes staggered along the $c$ cubic axis,
occurs below the N\'eel temperature $T_N\simeq 140$ K \cite{Kov10}.
The purpose of this paper is to provide a comprehensive and unbiased
scenario of these two phase transitions starting from the spin-orbital
model of Ref. \cite{Fei99} using the realistic parameters.

Up to now, to the best of our knowledge, the phase transitions in
LaMnO$_3$ were investigated only using classical on-site mean field (MF)
theory which factorizes intersite spin and orbital correlations. Indeed,
the $A$-AF phase with FM planes was obtained for LaMnO$_3$
\cite{Fei99}, and the orbital excitations were described within the
orbital model and investigated at zero temperature \cite{vdB99}.
Unfortunately, no thermal analysis of the spin-orbital model for
LaMnO$_3$ was performed so far beyond the early MF study \cite{Fei99},
where the importance of orbital interactions induced by JT distortions
was pointed out. This result is confirmed by calculations of the
transition temperature from the superexchange mechanism by using the
local density approximation combined with dynamical mean field method
\cite{Pav10}. Below we provide evidence that going beyond the on-site MF
theory is necessary to get a physically correct insight into the onset
of $C$-type AO order below the orbital phase transition and its changes
below the magnetic one. We analyze the nature of orbital states
occupied by $e_g$ electrons in LaMnO$_3$.

The main purpose of this work is to carry out a systematic and
consistent analysis of the effective spin-orbital model for LaMnO$_3$
within the cluster MF approach. In particular, of primary importance
is to establish the orbital order by working out the value of orbital
mixing angle (and its temperature dependence). The second objective of
this work is to evaluate spin-orbital entanglement --- both for
(i) on-site and
(ii) on-bond correlations.
As we show below the obtained results justify \textit{a posteriori}
the commonly used spin-orbital decoupling approximation. The third
objective is to establish the range for the JT coupling constant.
The complexity of the model gives rise to technical benchmark.
The technical objective is to verify to what extent the cluster
calculations are reliable.

The paper is organized as follows: In Sec. \ref{sec:som} we introduce
the spin-orbital model for LaMnO$_3$ and present a detailed discussion
of various terms contributing to superexchange in Sec. \ref{sec:HJ}.
The JT terms and parameters are presented in Sec. \ref{sec:HJT}.
Next we introduce the essential features of a MF analysis
of spin and orbital order in Sec. \ref{sec:MF} and discuss the main
problems related to the analysis of coexisting $A$-AF and $C$-AO order
in Sec. \ref{sec:preli}. Cluster MF approaches are presented in Sec.
\ref{sec:CMF}. In Sec. \ref{sec:resu} we present the numerical results
obtained using different clusters with disentangled interactions and
also entangled spin-orbital bond for magnetic and orbital transition,
see Sec. \ref{sec:phd}. Next we give spin and orbital bond correlations
and exchange constants, see Secs. \ref{sec:spin}-\ref{sec:orbi}, as
well as optical spectral weights, see Sec. \ref{sec:sw}. In Sec.
\ref{sec:enta} we investigate on-site (Sec. \ref{sec:site}) and on-bond
(Sec. \ref{sec:bond}) spins-orbital entanglement and finally its impact
on the value of N\'eel transition temperature, see Sec. \ref{sec:TN}.
This analysis allows us to formulate general conclusions concerning
the role of entanglement in the properties of LaMnO$_3$.
The summary and conclusions are given in Sec. \ref{sec:summa}.

\section{Model}
\label{sec:som}

The model analyzed in this work consists of two terms:
(i) the superexchange interaction $H_J$ derived from
$d_i^4d_j^4 \rightleftharpoons d_i^5d_j^3$ charge excitations on
bonds between nearest neighbor Mn$^{3+}$ ions and
(ii) the JT term $H_{\rm JT}$ which follows from local lattice
distortions triggered by the JT coupling. The Hamiltonian reads,
\begin{equation}
 \label{eq:model}
{\cal H} = H_J +  \mathbf{1}_{\mbox{spin}} \otimes H_{\rm JT} .
\end{equation}
The superexchange part $H_J$ was derived by considering charge
excitations for both $e_g$ and $t_{2g}$ electrons \cite{Fei99}, where
it was shown that it predicts $A$-AF order for the realistic parameters.
Although the charge excitations on oxygen orbitals play a role
\cite{Ole05}, its generic form which includes such excitations only in
an effective way was successfully used to interpret the temperature
evolution of spectral weights measured in the optical spectroscopy
\cite{Kov10}. The model Eq. (\ref{eq:model}) poses a difficult many-body
problem. We treat it here in several approximations discussed below and
investigate to what extent spins and orbitals may be separated from each
other.

\subsection{Orbital projections}
\label{sec:proo}

To establish formulas for $H_J$ and $H_{\rm JT}$ one has to introduce
first orbital projection operators for $e_g$ orbital degree of freedom.
They are the same as for the Kugel-Khomskii model for KCuF$_3$
\cite{Ole00}.
We start with on-site orbital projection operators at site $i$:
\begin{equation}
 P_i^{\zeta_\gamma}\equiv |i\zeta_\gamma\rangle\langle i\zeta_\gamma|, \qquad
 P_i^{\xi_\gamma}  \equiv |i\xi_\gamma  \rangle\langle i\xi_\gamma|.
\end{equation}
Here the orbitals $\{|\zeta_{\gamma}\rangle,|\xi_{\gamma}\rangle\}$
form an orthogonal basis at site $i$ consisting of a directional orbital
$|\zeta_{\gamma}\rangle$ and a planar orbital $|\xi_{\gamma}\rangle$.
They depend on the cubic axis index $\gamma$:
\begin{center}
\begin{tabular}{C{3cm}C{3cm}}
$|\zeta_a\rangle= |3x^2 -r^2\rangle$, &$|\xi_a\rangle= |y^2-z^2\rangle$, \\
$|\zeta_b\rangle= |3y^2 -r^2\rangle$, &$|\xi_b\rangle= |z^2-x^2\rangle$, \\
$|\zeta_c\rangle= |3z^2 -r^2\rangle$, &$|\xi_c\rangle= |x^2-y^2\rangle$.
\end{tabular}
\end{center}
The projection operators exhaust the orbital space, i.e.,
\begin{equation}
 P_{i}^{\zeta_\gamma}+P_{i}^{\xi_\gamma}=1.
\end{equation}

Depending on the bond direction on which the superexchange $J_J$ or JT
interaction $H_{\rm JT}$ acts, one has to choose $\gamma=a,b,c$.
Then, it is necessary to introduce projection operators for a bond
along the cubic axis $\gamma$:
\begin{align}
 \mathcal{P}^{\zeta\xi}_{\langle ij \rangle\parallel\gamma}
 &\equiv
 2P_{(i}^{\zeta_\gamma}P_{j)}^{\xi_\gamma} \equiv
 P_{i}^{\zeta_\gamma}P_{j}^{\xi_\gamma}+P_{i}^{\xi_\gamma}P_{j}^{\zeta_\gamma},
\label{zetxi}
 \\
 \mathcal{P}^{\zeta\zeta}_{\langle ij \rangle\parallel\gamma}
 &\equiv
 2 P_{i}^{\zeta_\gamma} P_{j}^{\zeta_\gamma},
\label{zetzet}
 \\
 \mathcal{P}^{\xi\xi}_{\langle ij \rangle\parallel\gamma}
 &\equiv
 2 P_{i}^{\xi_\gamma} P_{j}^{\xi_\gamma}.
\label{xixi}
\end{align}
Note that the operator $2P_{(i}^{\zeta_\gamma}P_{j)}^{\xi_\gamma}$
refers to a symmetrized product of two projection operators on sites
$i$ and $j$. Such operators characterize the processes which
contribute to various interaction terms, see below. The factor of 2 in
Eqs. (\ref{zetzet}) and (\ref{xixi}) is introduced for a more compact
notation below as the charge excitations used to derive the
superexchange involve intersite excitations by hopping which couples
two directional $\{|i\zeta_{\gamma}\rangle,|j\zeta_{\gamma}\rangle\}$
orbitals at sites $i$ and $j$ only \cite{Ole00}. Actually, the bond
operators Eqs. (\ref{zetxi})-(\ref{xixi}) are not independent and obey
the constraint which may serve to determine one of them from the other
two,
\begin{equation}
\mathcal{P}^{\zeta\zeta}_{\langle ij \rangle}
 + 2\mathcal{P}^{\zeta\xi}_{\langle ij \rangle}
 + \mathcal{P}^{\xi\xi}_{\langle ij \rangle} = 2.
\end{equation}

\subsection{Spin-orbital superexchange}
\label{sec:HJ}

The superexchange part $H_J$ of the effective Hamiltonian ${\cal H}$
(\ref{eq:model}) arises from various virtual excitations to Mn$^{2+}$
states with $d^5$ electronic configuration \cite{Fei99}. The charge
excitations $d_i^4d_j^4\rightleftharpoons d_i^5d_j^3$ obtained from
$e_g$ electron hopping may generate a high-spin (HS) $^6A_1$ state or
low-spin (LS) $^4A_1$, $^4E$ and $^4A_2$ states. The remaining Mn$^{4+}$
ion is always in HS $^4A_1$ state. In addition, there are also charge
excitations by a transfer of a $t_{2g}$ electron between sites $i$ and
$j$. The $t_{2g}$ levels are half filled in the Mn$^{3+}$ $d^4$ ground
state, so only LS excited states are generated by the $t_{2g}$ electron
hopping as the orbital flavor is again conserved. We list here the
excited $d_i^5d_j^3$ states pairwise for $d_i^5d_j^3$: $(^4T_1,^4T_1)$,
$(^4T_1,^4T_2)$, $(^4T_2,^4T_1)$, and $(^4T_2,^4T_2)$.

The leading part of the superexchange Hamiltonian $H_J$ describes the
interactions obtained from charge excitations by $e_g$ electrons and
is parameterized by four physical parameters:
(i) hopping element for $e_g$ electrons $t$ between two $|\zeta\rangle$
orbitals oriented along the considered bond $\langle ij\rangle$,
(ii) on-site Coulomb repulsion energy~$U$,
(iii) Hund's exchange $J_H^e$ for $e_{g}$ electrons, and
(iv) Hund's exchange $J_H^t$ for $t_{2g}$ electrons.
In fact, $U$ and $J_H^e$ may be expressed by three Racah parameters
\cite{Ole05}: $U=A+4B+3C$, $J_H^e=4B+C$.
For charge excitations generated by $t_{2g}$ electrons Hund's exchange
is somewhat smaller, $J_H^t=3B+C$, and the hopping for a $\pi$ bond
is reduced to $\beta t$, where $\beta=1/9$.
All the excitation energies are collected in Table \ref{tab:exci}.

\begin{table}[t!]
\caption{The $d_i^5$ excited states and their excitation energies $E_n$
found from the analysis of $d_i^4d_j^4\leftrightarrows d_i^5d_j^3$ charge
excitations in undoped LaMnO$_3$.
Parameters to calculate $E_n$ as in Eqs. (\ref{para}) and (\ref{kappa}).
The experimental energies $E_{\textrm{exp}}$ are from Ref. \cite{Kov10}.}
\begin{ruledtabular}
 \begin{tabular}{cclcc}
  &     & \multicolumn{3}{c}{excitation energy $E_n$}  \\
 orbital &     term        & $E_n$ (theory) & $E_n$ (eV) &$E_{\textrm{exp}}$ (eV)\\
\hline
  $e_g$  &   $^6A_1$       & $U- 3J_H^e$    &    1.93    &     $2.0\pm0.1$       \\
         &   $^4A_1$       & $U+ 3J_H^e/4$  &    4.52    &     $4.3\pm0.2$       \\
         &   $^4E$         & $U+ 5J_H^e/4$  &    4.86    &     $4.6\pm0.2$       \\
         &   $^4A_2$       & $U+13J_H^e/4$  &    6.24    &     $6.1\pm0.2$       \\
\hline
$t_{2g}$ & $(^4T_1,^4T_2)$ & $U+ 5J_H^t/4$  &    4.74    &     $3.9\pm0.1$       \\
         & $(^4T_2,^4T_2)$ & $U+ 9J_H^t/4$  &    5.33    &     $4.4\pm0.1$       \\
         & $(^4T_1,^4T_1)$ & $U+11J_H^t/4$  &    5.62    &     $4.8\pm0.1$       \\
         & $(^4T_2,^4T_1)$ & $U+15J_H^t/4$  &    6.21    &     $5.7\pm0.5$       \\
 \end{tabular}
\end{ruledtabular}
\label{tab:exci}
\end{table}

Whereas there are four physical parameters that control the Hamiltonian
$H_J$, the model has three independent parameters. The superexchange
constant is \cite{Fei99},
\begin{equation}
 J = \frac{4t^2}{U},
\end{equation}
where $t$ is the hopping element for a bond $\langle ij\langle$
(via oxygen ions) between two directional $e_g$ orbitals
$\{|\zeta_{i\gamma}\rangle,|\zeta_{j\gamma}\rangle\}$, e.g.
$|3z^2-r^2\rangle$ orbitals along the $c$ axis.
To specify various contributions we introduce
\begin{equation}
 \eta_e=\frac{J_H^e}{U}, \hskip .7cm
 \eta_t=\frac{J_H^t}{U} .
\end{equation}
to distinguish between Hund's exchange for $e_g$ and $t_{2g}$
electrons. This generates small changes of the
superexchange Anderson contributions to $H_J$ considered before in
Refs. \cite{Ole05} and \cite{Kov10}. The Goodenough contributions which
stem from charge excitations on oxygen orbitals along Mn-O-Mn bonds
\cite{Ole05} are neglected here in the effective model $H_J$.

Now, it it straightforward to write the formula for $H_J$ which
includes the terms due to $e_g$ ($H_e$) and $t_{2g}$ ($H_t$) electron
excitations,
\begin{equation}
 H_J = H_e + H_t,
\label{HJ}
\end{equation}
where
\begin{align}
\label{He}
\begin{split}
 H_e = J\sum_{\langle ij \rangle\parallel\gamma}
&\Bigg\{-\frac{1}{40} r_1
      \big( \vec S_i \cdot \vec S_j + 6 \big) \mathcal{P}^{\zeta\xi}_{\langle ij \rangle}    \\
    &+\left[ \frac{1}{64} r_3 + \frac{3}{320} r_2 \right]
      \big( \vec S_i \cdot \vec S_j - 4 \big) \mathcal{P}^{\zeta\xi}_{\langle ij \rangle}    \\
    &+\left[ \frac{1}{64} r_4 + \frac{1}{64} r_5 \right]
      \big( \vec S_i \cdot \vec S_j - 4 \big) \mathcal{P}^{\zeta\zeta}_{\langle ij \rangle}
\Bigg\},
\end{split}
\end{align}
and
\begin{equation}
\label{Ht}
 H_t = J \beta  \sum_{\langle ij \rangle}
       \frac{1}{8} r_t
       \big( \vec S_i \cdot \vec S_j - 4 \big).
\end{equation}
The multiplet structure of $e_g$ excited states, see Table
\ref{tab:exci}, is given by
\begin{eqnarray}
 r_1&=& \frac{1}{1-3\eta_e}, \hskip 1.5cm
 r_2 = \frac{1}{1+3\eta_e/4},  \nonumber \\
 r_3&=& r_4 = \frac{1}{1+5\eta_e/4}, \qquad
 r_5 = \frac{1}{1+13\eta_e/4}.
\end{eqnarray}
For excitations by a $t_{2g}$ electron we have to collect all the terms
by projecting the excited configurations onto the eigenstates listed in
Table \ref{tab:exci}. This results in a single coefficient,
\begin{equation}
r_t\! =\!
 \frac{1}{4}\left[ \frac{1}{1+5\eta_t/4}\!+\! \frac{1}{1+9\eta_t/4}
\!+\!\frac{1}{1+11\eta_t/4}\!+\! \frac{1}{1+15\eta_t/4} \right].
\end{equation}

\subsection{Jahn-Teller terms and parameters}
\label{sec:HJT}

The JT Hamiltonian $H_{\rm JT}$ describes the coupling between the
adjacent sites via the mutual octahedron distortion. We invoke
here the simple quantum 120\textdegree~model to control it. We note
that JT effect is connected with the oxygen atoms displacements which
result in longer and shorter Mn-O bonds but leave the manganese
positions undisturbed.

The correct form of the effective interactions between the occupied
$e_g$ orbitals depends only on the symmetries of the Mn-O octahedra.
A classical derivation of the effective orbital-orbital interactions
between the neighboring manganese ions was presented by Halperin and
Englman \cite{Hal71}. After transforming these interactions to the
bond orbital projection operators Eqs. (\ref{zetxi})-(\ref{xixi}),
the formula that describes $H_{\rm JT}$ reads as follows \cite{Fei99}:
\begin{equation}
 H_{\rm JT} = \kappa
          \sum_{\langle ij \rangle }
          \left(   \mathcal{P}^{\zeta\zeta}_{\langle ij \rangle}
                - 2\mathcal{P}^{\zeta\xi}_{\langle ij \rangle}
                + \mathcal{P}^{\xi\xi}_{\langle ij \rangle} \right).
\label{HJT}
\end{equation}
We remark that a full derivation of the orbital-orbital interaction
(\ref{HJT}) generated by the JT coupling is complicated \cite{Geh75}
and here we limit ourselves to presenting the consequences of its
simplified form (\ref{HJT}) that follows from the classical
derivation \cite{Hal71}. As a result,
the JT term is controlled by a single parameter $\kappa$ that describes
the rigidity of the oxygen positions (or magnitude of displacement
caused by the adjacent Mn orbital state). If the oxygens were rigid and
their positions were not influenced by manganese orbital states then
$\kappa=0$, but in reality $\kappa>0$.

The choice of parameters which determine the superexchange has been
extensively discussed in the past, for instance in Ref. \cite{Kov10}.
Here we use the following values (all in eV):
\begin{equation}
 t=0.37, \quad U = 4.0, \quad J_H^e=0.69, \quad J_H^t=0.59,
\label{para}
\end{equation}
to obtain the excitation energies given in Table \ref{tab:exci}. These
values stay in reasonable agreement with the experimental values that
are also listed in Table  \ref{tab:exci}. In  addition, we adopt here
\begin{equation}
 \kappa = 6\ {\rm meV}.
\label{kappa}
\end{equation}
This value was chosen \textit{a posteriori} to fit the value of the
orbital transition temperature $T_{\rm OO}$. It confirms the earlier
observation \cite{Oka02} that the JT coupling is just a fraction of
$J=t^2/U\simeq 34$ meV, i.e., it is much smaller than the values
suggested for $\kappa$ within the MF approach at the early stage of
the theory, such as: $11$ meV \cite{Fei99} and $9.1$ meV \cite{Sik03}.

\section{Mean field treatment and beyond}
\label{sec:MF}

\subsection{Orbital state description}
\label{sec:omf}

The simplest approach to the many-body problem posed by Eq.
(\ref{eq:model}) is the MF approach where only single-site averages are
introduced to investigate possible order. In spin space we consider the
symmetry breaking along the $z$ axis, with $\langle S_i^z\rangle$ being
the order parameter at site $i$. In a uniform phase the parameter
$\langle S^z\rangle\equiv\langle S_i^z\rangle$ is a magnetic (spin)
order parameter.

The orbital (pseudospin) state is completely described by density matrix
\begin{equation}
  \rho
  \equiv
  \operatorname{Tr}' \Big( | \Psi \rangle \langle \Psi | \Big)
  \in \mathbb{C}^{2\times2}
\end{equation}
where $\Psi$ denotes the complete wave function and
$\operatorname{Tr}'$ means that all but considered pseudospin degrees
of freedom were integrated out. When the orbital basis,
\begin{eqnarray}
| 3z^2-r^2\rangle/\sqrt{6}\,&\Rightarrow &
\begin{pmatrix} 1 \\ 0 \end{pmatrix} , \\
| x^2-y^2 \rangle/\sqrt{2}\,&\Rightarrow &
\begin{pmatrix} 0 \\ 1 \end{pmatrix},
\end{eqnarray}
is chosen, the density operator $\rho$ is isomorphic to $2\times2$ not
negative-definite hermitian matrix with trace equal to one.
For the sake of description of orbital (pseudospin) Bloch vector was
invoked, i.e., vector $\vec r \in \mathbb{R}^2$ contained in closed
unit disk for which
\begin{equation}
  \rho =\frac{1}{2} \mathbf{1}_2 + \frac{1}{2}\vec r \cdot\vec\sigma,
  \qquad
  \text{where: }
    \vec \sigma =   \begin{pmatrix} \sigma_x \\ \sigma_z \end{pmatrix} .
\end{equation}
Then, the radius $r$ and the angle $\theta$ are introduced
to replace the polar coordinates,
\begin{equation}
\vec r = r\begin{pmatrix} +\sin\theta \\ -\cos\theta \end{pmatrix}, \quad
  \begin{array}{c}
   r \in [0,1], \\
   \theta \in (-180\text{\textdegree} ,180\text{\textdegree}].
  \end{array}
\end{equation}
In terms of $r$ and $\theta$ parameters one finds:
 \begin{align}
  \langle\sigma_x\rangle &= \text{Tr}(\sigma_x\rho) = +r \sin\theta,\\
  \langle\sigma_z\rangle &= \text{Tr}(\sigma_z\rho) = -r \cos\theta.
 \end{align}
Note, that the pure states lie on the unit circle ($r=1$),
and the corresponding wave vectors are equal to  \cite{vdB99},
\begin{equation}
| \theta \rangle=
  \cos(\theta/2) | 3z^2-r^2\rangle/\sqrt{6}
+ \sin(\theta/2) | x^2-y^2 \rangle/\sqrt{2}.
\label{theta}
\end{equation}
The amplitude $r$ is considered as orbital order parameter.

\subsection{Preliminary analysis of $A$-AF/$C$-AO order}
\label{sec:preli}

At low temperature $A$-AF and $C$-AO order coexist in LaMnO$_3$.
Let us start the analysis of the symmetry broken state with the FM
$(a,b)$ planes (perpendicular to the cubic $c$ axis). In this case
$\langle\vec S_i\cdot\vec S_j\rangle=+4$ for the classical bonds in
the plane, and the only nonvanishing contribution from the
superexchange Hamiltonian is the interaction involving the HS states.
We observe that this part of Hamiltonian (\ref{He}) is proportional
to $\mathcal{P}^{\zeta\xi}_{\langle ij\rangle}$, and its coefficient
is approximately equal to $-70$ meV. The JT effect enhances this
interaction further.
The interaction parallel to the $a$ axis prefers the pair of orbitals
proportional to $|x^2\rangle$ and $|z^2-y^2\rangle$, respectively
(with the mixing angle $\theta$ equal to $-120$\textdegree\ and
$+60$\textdegree). At the same time
the interaction parallel to the $b$ axis prefers the pair of orbitals
proportional to $|y^2\rangle$ and $|z^2-x^2\rangle$ respectively
(with the mixing angle $\theta$ equal to $+120$\textdegree\ and
$-60$\textdegree). These interactions are frustrated. As a result of
them the ground state emerges which involves AO order for the orbitals
with mixing angle $\theta$ alternating between 90\textdegree\ and
$-90$\textdegree.

In the AF direction for the bonds along the $c$ axis the problem is
more subtle. In this case $\langle\vec S_i\cdot\vec S_j\rangle=-4$ in
MF (i.e., neglecting quantum fluctuations) and superexchange terms for
both HS and LS excitations are important. They are found in $H_J$ both
as proportional to $\mathcal{P}^{\zeta\xi}_{\langle ij\rangle}$, and
proportional to $\mathcal{P}^{\zeta\zeta}_{\langle ij\rangle}$.
The former term proportional to $\mathcal{P}^{\zeta\xi}$ contributes
with coefficient $\approx -35$ meV, while the JT effect enhances this
interaction. This term favors AO order with $3z^2-r^2$ ($\theta=0$)
and $x^2-y^2$ ($\theta=180$\textdegree) orbital pairs. But it is not
strong enough and the pattern with $\theta$ equal to
$\pm90$\textdegree\ sustain. (In fact, this configuration is
energetically not very bad for this interaction).
In addition the term proportional to
$\mathcal{P}^{\zeta\zeta}_{\langle ij\rangle}$ is present with negative
coefficient. This interaction enhances the $|3z^2-r^2\rangle$
($\theta=0$) contributions to the orbitals in the ground state. Due to
these interactions the mixing angle $\theta$ is significantly lower
than 90\textdegree.

We may check our tentative results against the crystallographic data
\cite{Huang97}. We invoke here the results for polymorphic structure
with symmetry $Pmna$ and lattice parameters: $a=5.7$\AA{}, $b=7.7$\AA{},
and $c=5.5$\AA{}. The AO pattern in the FM planes should result in
the alternating pattern of the octahedron distortions. Indeed, in real
crystal such a distortion pattern was observed.
The in-plane Mn-O bonds length are equal to $s=1.91$\AA{}, $l=2.18$\AA{}.
What is more the Mn-O bonds perpendicular to the FM planes are
longer than 1.91\AA{} (their length is equal to $m=1.96$\AA{}).
It means that the alternating mixing angle $\theta$ is lower than
120\textdegree.

One may express the distortion in terms of $Q_2$ and $Q_3$ modes.
The corresponding coefficients are equal to
\begin{equation}
 Q_2 = \frac{l-s}{\sqrt{2}}, \qquad
 Q_3 = \frac{2m-l-s}{\sqrt{6}}.
\end{equation}
From $(Q_2,Q_3)$ one may evaluate the orbital mixing angle with the aid
of identification
\begin{equation}
 \big(Q_2, Q_3\big) = |Q|\big(\sin\theta, \cos\theta\big) .
\end{equation}
This leads to $\theta=108$\textdegree.
We see that in real crystal $\theta$ is bigger than 90\textdegree;
in contrast for our model where $\theta\leq 90$\textdegree.

We may guess two possible reasons for this discrepancy.
\begin{itemize}
\item We should note that in a crystal of LaMnO$_3$ not only JT effect
is important. The JT effect is responsible for octahedron deformation
but leaves the Mn positions undisturbed. When the adjacent Mn-Mn
distances are not equal to each other the crystal-field interactions
occur. In case of the analyzed crystal the adjacent Mn-Mn distances are
equal for the Mn pairs belonging to one FM plane, but this distance
is larger than the distance between adjacent Mn from two different
FM planes. This means that there is one more interaction that plays
a role. In this case this interaction prefers instead $|x^2-y^2\rangle$
orbitals ($\theta=180$\textdegree) and may result in boosting the value
of the mixing angle. We do not consider this interaction within the
present work.

\item Due to anharmonicity of the JT interactions the potential
achieves its minimum in $\theta=0$\textdegree, $\pm120$\textdegree.
This may result in boosting the mixing angle value
(up to $120$\textdegree).
\end{itemize}

\section{Cluster mean field approaches}
\label{sec:CMF}

\subsection{Beyond single-site mean field theory}
\label{sec:so}

In this work the model described by Eq. (\ref{eq:model}) is analyzed
to figure out the orbital and spin temperature dependence, as well as
the phase transition temperatures, $T_{\rm N}$ and $T_{\rm OO}$. We
employ the cluster MF approach to investigate the interplay between
these phase transitions in  a more realistic way than in a single-site
MF theory. As in previous modern applications of the cluster MF method
\cite{Brz12,Alb11,Got16}, the considered cluster is embedded by the MF
terms of its surrounding, and self-consistent conditions are imposed
(to ensure that all the on-site mean values of surrounding vertices
take their proper values). In presented calculations $A$-AF and $C$-AO
order is assumed. It may be taken for granted as indeed this type of
order gives the lowest energy in the ground state at $T=0$.

The smallest cluster that captures the point symmetry of considered
Hamiltonian Eq. (\ref{eq:model}) (a chiral octahedral symmetry
$\text{O}_\text{h}$) consists of 7~sites.
Due to the big number of spin-orbital degrees of freedom per site
($5\times 2=10$) the corresponding space of states of the cluster is
enormous ($10^7$). This makes it impossible to carry out exact
(cluster) calculations. Thus, further approximations cannot be
avoided. In this work two alternative further approximation
(or calculation) schemes are invoked:
(i) \textit{multi-single-bond scheme} and
(ii) spin-orbital \textit{decoupling scheme}.

\subsection{Multi-single-bond calculation scheme}
\label{sec:multi}

First calculation scheme, called here \textit{multi-single-bond scheme},
is rooted in a single-bond calculation. As a first step a single
calculation for one bond is performed to find the state of this
bond immersed in its fixed environment. As the cubic directions are
nonequivalent in symmetry-broken states, in elementary iteration of
\textit{multi-single-bond scheme} 6~single-bond calculations are
carried out as there are 6 inequivalent (but symmetry related) bond
types in the considered system. At the end of each elementary iteration
all important single-site operator mean values are established as
averages of mean values achieved from corresponding single-bond
calculations (involving considered site).
Iterations are repeated until convergence.

The main advantage of the \textit{multi-single-bond scheme} is that it
treats all the degrees of freedom on equal footing, including the
spin-spin, orbital-orbital and spin-orbital interactions.
The main disadvantage stems from the small cluster size --- all the
phase transition temperatures predicted by it are considerably
overestimated.

Although \textit{multi-single-bond calculation scheme} suffers from
many problems it can provide some useful information. Firstly, with the
aid of it we can evaluate mean values: $\langle S_i^z\sigma_i^z\rangle$,
$\langle S_i^zS_j^z\sigma_i^z\sigma_j^z\rangle$, \textit{etcetera}.
Second, what is even more important, it can provide heuristic value of
the error that occurs in calculations in which the spin and the orbital
degrees of freedom are treated separately. To get appropriate error
values one should compare result from this model with the result of
the factorized model, i.e., modified in such away that all spin-orbital
interactions are artificially broken into sums of two product operators
and are disentangled.

\subsection{Decoupling calculation scheme}

Second calculation scheme, called here \textit{decoupling scheme} is
rooted in a decoupling of the spin and the orbital degree of freedom.
In this approach spin and orbital degrees of freedom are disentangled.
In the elementary iteration step of the \textit{decoupling scheme} one
pure spin calculation and one pure orbital calculation are carried out.
The pure orbital calculations Hamiltonian arises from Hamiltonian
Eq. (\ref{eq:model}) after replacement of all the products of spin
operators by the averages $\langle\vec S_i\cdot\vec S_j\rangle$ being
scalar parameters (one parameter for one bond direction). Similarly,
the pure spin calculations Hamiltonian arises from Hamiltonian
Eq. (\ref{eq:model}) after replacement of all orbital projection
operators $\mathcal{P}^{\zeta\xi}_{\langle ij\rangle\parallel\gamma}$
and $\mathcal{P}^{\zeta\zeta}_{\langle ij\rangle\parallel\gamma}$ by
the scalar parameters. After such a replacement one ends up with the
anisotropic spin model with exchange parameters $J_{ab}$ and $J_c$
for $S=2$ spins,
\begin{equation}
H_S = J_{ab}\sum_{\langle ij\rangle\parallel ab} \vec S_j\cdot\vec S_j
+ J_c\sum_{\langle ij\rangle\parallel c} \vec S_j\cdot\vec S_j.
\end{equation}
Here we use a short-hand notation and label by $J_{ab}$ exchange
interactions in $(a,b)$ planes, $J_a=J_b$, as the bonds along the $a$
and $b$ cubic axes are equivalent. In this case the $J_\gamma$'s
constants are given by
\begin{multline}
\label{J}
 J_\gamma =
 J\Bigg\{
   \left[ -\frac{1}{40} r_1 + \frac{1}{64} r_3 + \frac{3}{320} r_2 \right]
\left\langle\mathcal{P}^{\zeta\xi}_{\langle ij\rangle}\right\rangle   \\
  +\left[ \frac{1}{64} r_4 + \frac{1}{64} r_5 \right]
\left\langle\mathcal{P}^{\zeta\zeta}_{\langle ij\rangle}\right\rangle
  + \beta \frac{1}{8} r_t
 \Bigg\},
\end{multline}
for $i$ and $j$ forming a bond $\langle ij\rangle\parallel\gamma$.

The pure spin Hamiltonian has some parameters which are connected with
the orbital state, and the pure orbital Hamiltonian has some parameters
which are connected with the spin state. At the end of each elementary
iteration of the present \textit{decoupling scheme}, the collection of
new parameters is established. The procedure is carried out until its
full convergence.

\begin{figure}[t!]
   \includegraphics[]{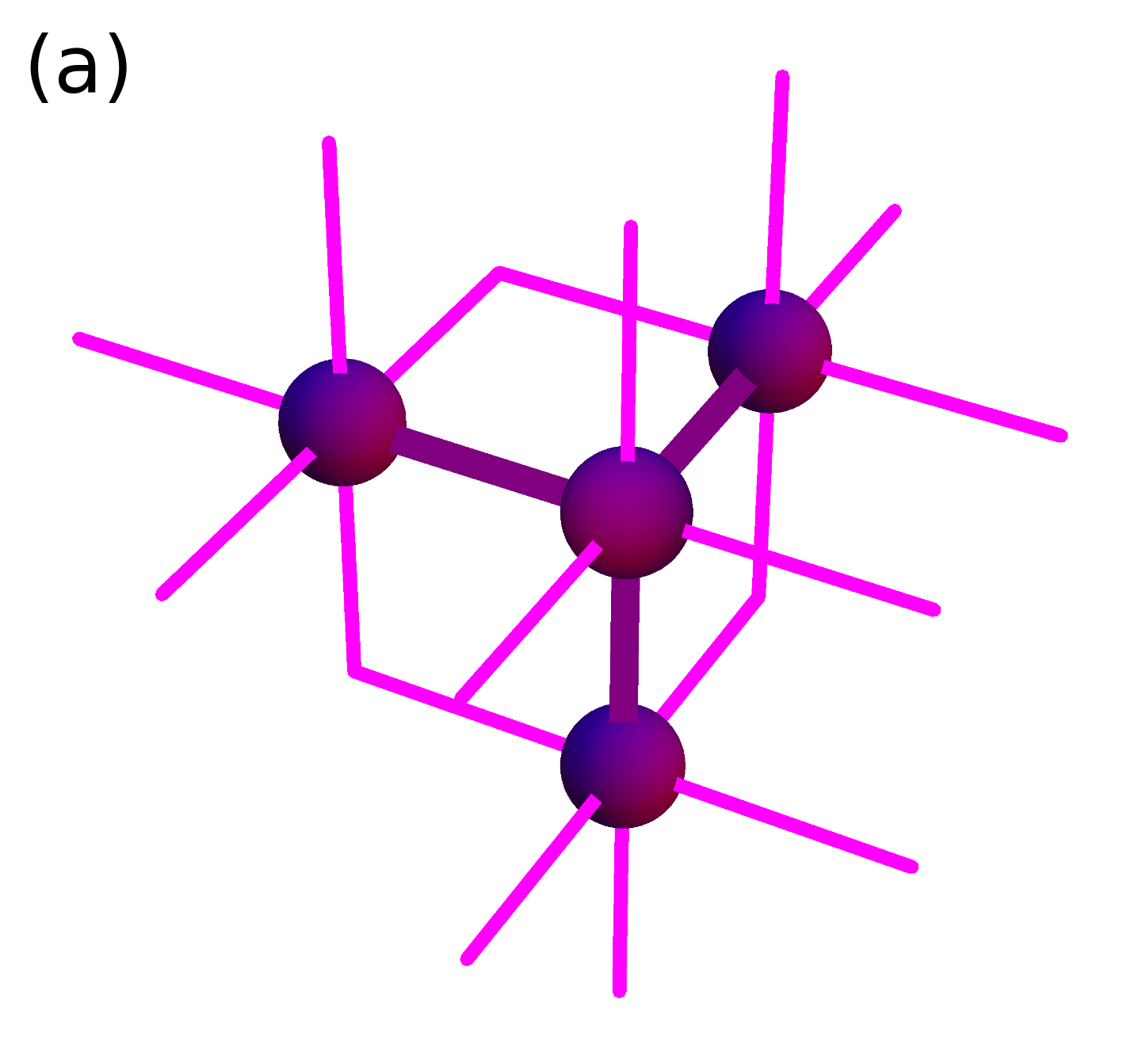}
   \includegraphics[]{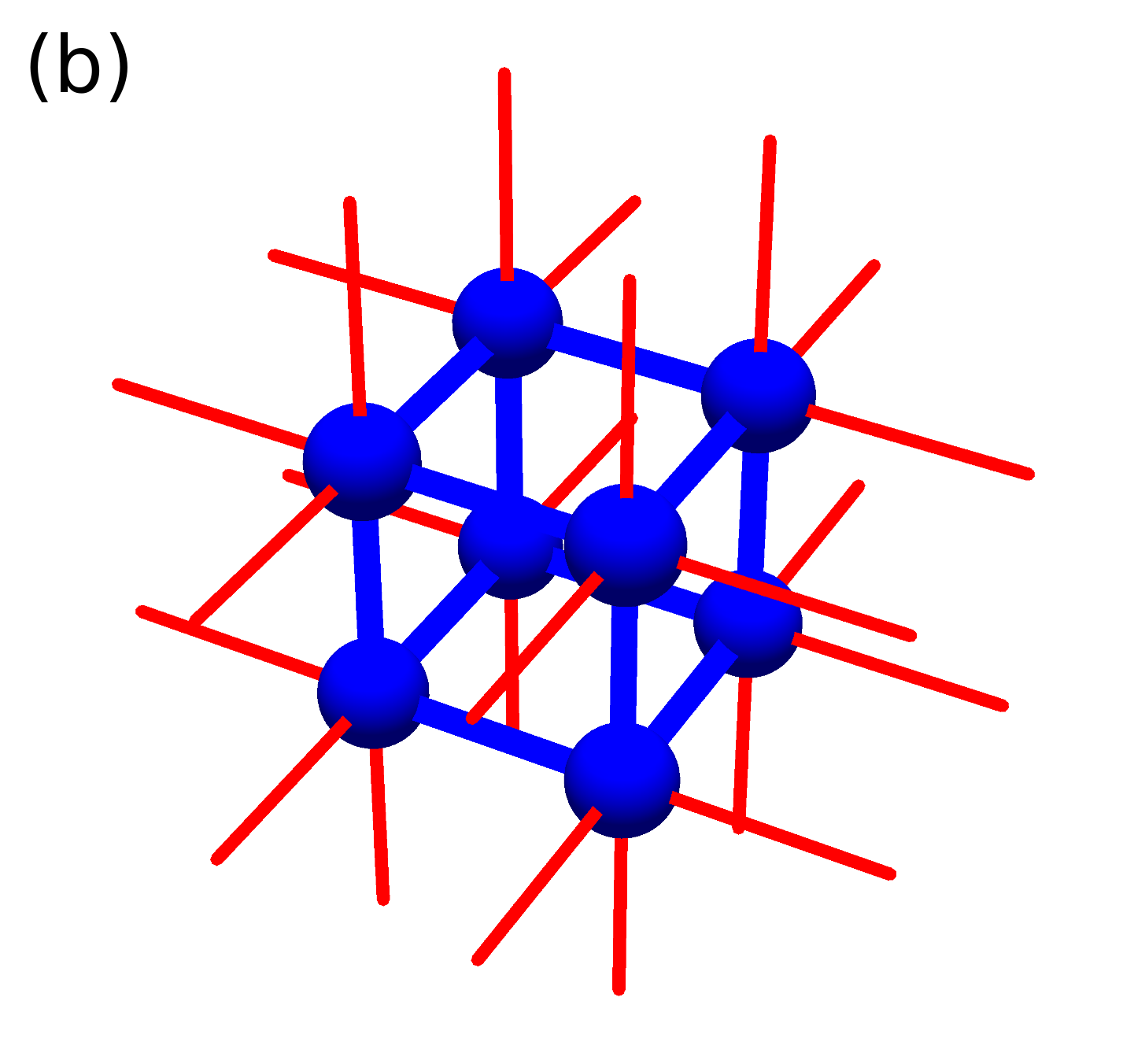}
   \includegraphics[]{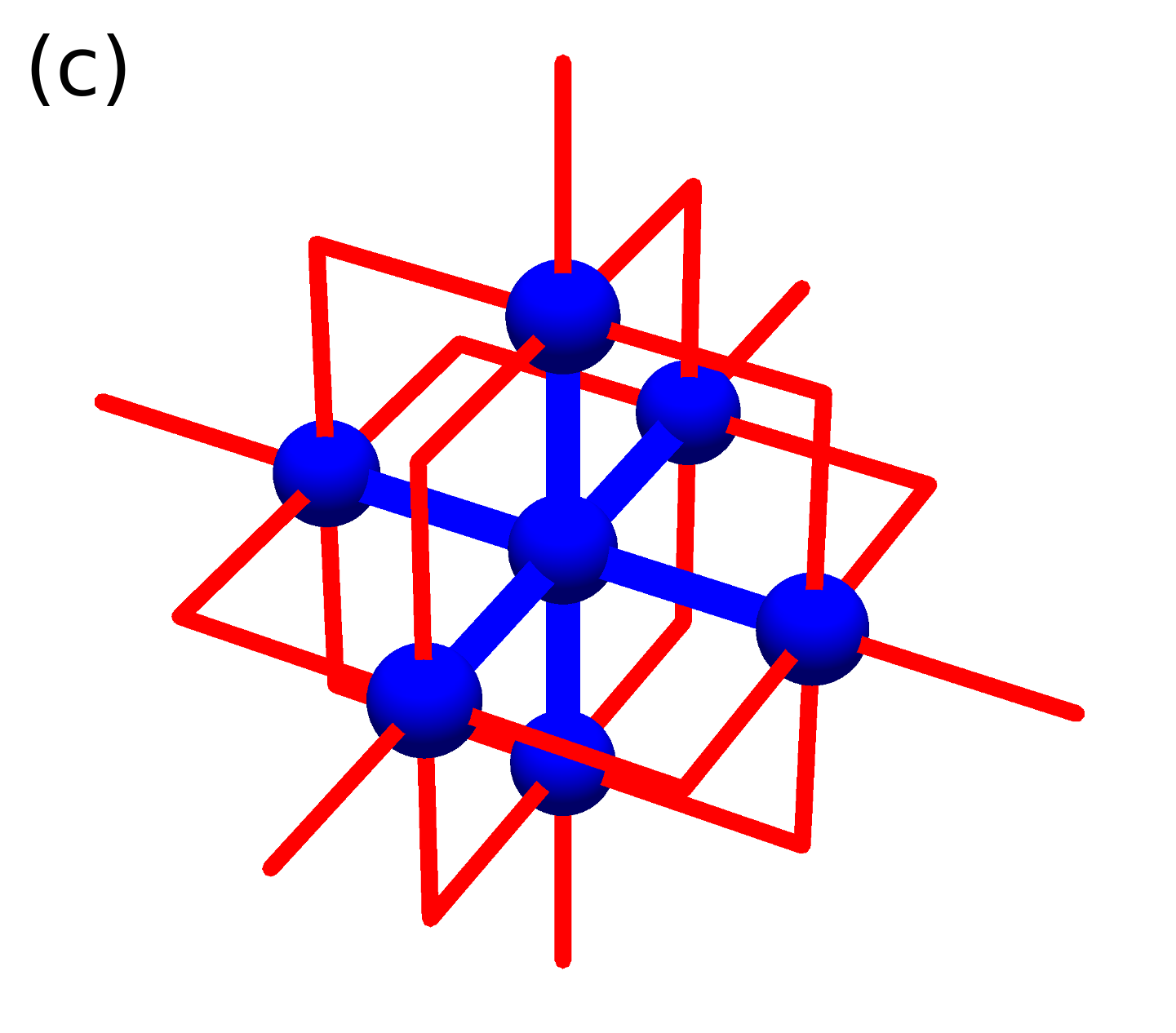}
\caption{The clusters used in the calculations:
(a) sputnik cluster for pure spin calculation, and
clusters for pure orbital calculations,
(b) cubic and
(c) hedgehog.
The vertices belonging to the cluster are depicted as spheres. The
thick segments that link the spheres represent the two-sites
interactions; the remaining lines represent interactions that
are considered in MF approximation.}
 \label{fig:clu}
\end{figure}

At the first glance is seems that method rooted in solely pure orbital
calculation would be valid for temperatures well above $T_{\rm N}$.
In fact, well above the onset of spin order at $T_{\rm N}$ the average
spin order parameter $\langle S^z\rangle$ is equal to zero and the spin
fluctuations on bonds seem to have marginal impact. Namely, there is a
temptation to build orbital model by changing the spin operators into
some given values and leave the orbital operators unchanged. This
simple concept is however misleading due to quite a complicated reason.
If one replaces both $\langle\vec S_i\cdot\vec S_j\rangle_{ab}$ and
$\langle\vec S_i\cdot\vec S_j\rangle_c$ by the same value (e.g. zero)
one creates the Hamiltonian that possesses additional symmetry.
The states derived from $C$-AO states by rotating by angle $+\phi$ all
the pseudospins in odd distinguished planes and by angle $-\phi$ in
even distinguished planes are degenerate. The system may gain energy by
lowering its symmetry by turning into a four-sublattice pattern.

Indeed, the calculations with
$\langle\vec S_i\cdot\vec S_j\rangle_{ab}=\langle\vec S_i\cdot\vec S_j\rangle_c$
predict that below a certain critical temperature the system develops
four-sublattice order. On the other hand, if one replaces
$\langle\vec S_i\cdot\vec S_j\rangle_{ab}$ by the some value and
$\langle\vec S_i\cdot\vec S_j\rangle_c$ by a different value, the
calculations predict that above $T_{\rm OO}$ the order parameter $r$
is still greater than zero. In reality
$\langle\vec S_i\cdot\vec S_j\rangle_{ab}\neq\langle\vec S_i\cdot\vec S_j\rangle_c$
below $T_{\rm OO}$ but
$\langle\vec S_i\cdot\vec S_j\rangle_{ab}-\langle\vec S_i\cdot\vec S_j\rangle_c\rightarrow 0$
as $T\rightarrow T_{OO}$ and
$\langle\vec S_i\cdot\vec S_j\rangle_{ab}=\langle\vec S_i\cdot\vec S_j\rangle_c\simeq 0$
for temperatures $T\geq T_{OO}$. By comparing these temperature regimes
we have established that there is no way to introduce one universal set
of $\langle\vec S_i\cdot\vec S_j\rangle_{ab}$ and
$\langle \vec S_i \cdot \vec S_j \rangle_c$ parameter values. So even
in high temperature calculations one has to bother about dynamical
generation of $\langle\vec S_i\cdot\vec S_j\rangle_{ab}$ and
$\langle\vec S_i\cdot\vec S_j\rangle_c$ values, i.e., one has to carry
out explicit calculations within the present \textit{decoupling scheme}.

The spin calculations was carried out on a cluster shown in Fig.
\ref{fig:clu}(a) (called here sputnik) and the orbital calculations
were carried out on clusters shown in Figs. \ref{fig:clu}(b) (called
here cubic) and \ref{fig:clu}(c) (called here hedgehog).
Note that in the cubic cluster all sites are treated on equal footing
--- all its sites have 3 bonds treated exactly (for orbitals only)
and 3 other MF bonds. In contrast, in the hegdehog cluster one site is
distinguished --- it has all 6 bonds treated exactly. In the latter
case all but distinguished site create the first coordination sphere,
which makes MF approximation less obscure (in point of view of the
distinguished site). In this sense we claim that the results
concerning this central site are more reliable than others.

The calculation scheme that was used to figure out the hedgehog cluster
state was quite different from the scheme used to figure out the cubic
cluster state. The difference originates from the fact that in the
cubic cluster all contained sites are treated on equal footing, whereas
in the hedgehog cluster one site is distinguished. The difference is in
implementing the MF of the cluster surroundings in each case. In the
cubic cluster case the environment is simply created as if it was the
collection of displaced cluster cubes. In the hedgehog cluster case we
take advantage of the existence of one distinguished site (which has no
MF interaction). Following this reasoning the surroundings site states
are assumed to be equal to the state of this distinguished site,
or this state with reversed sign of $\theta$ angle --- depending on
sublattice affiliation. The sputnik cluster has one distinguished site
and the calculations being carried out are similar in the operation to
the calculations concerning the hedgehog cluster.

\section{Numerical results and discussion}
\label{sec:resu}

\subsection{Order parameters and cluster dependence}
\label{sec:on-site}

\begin{figure}[t!]
   \includegraphics[width=\columnwidth]{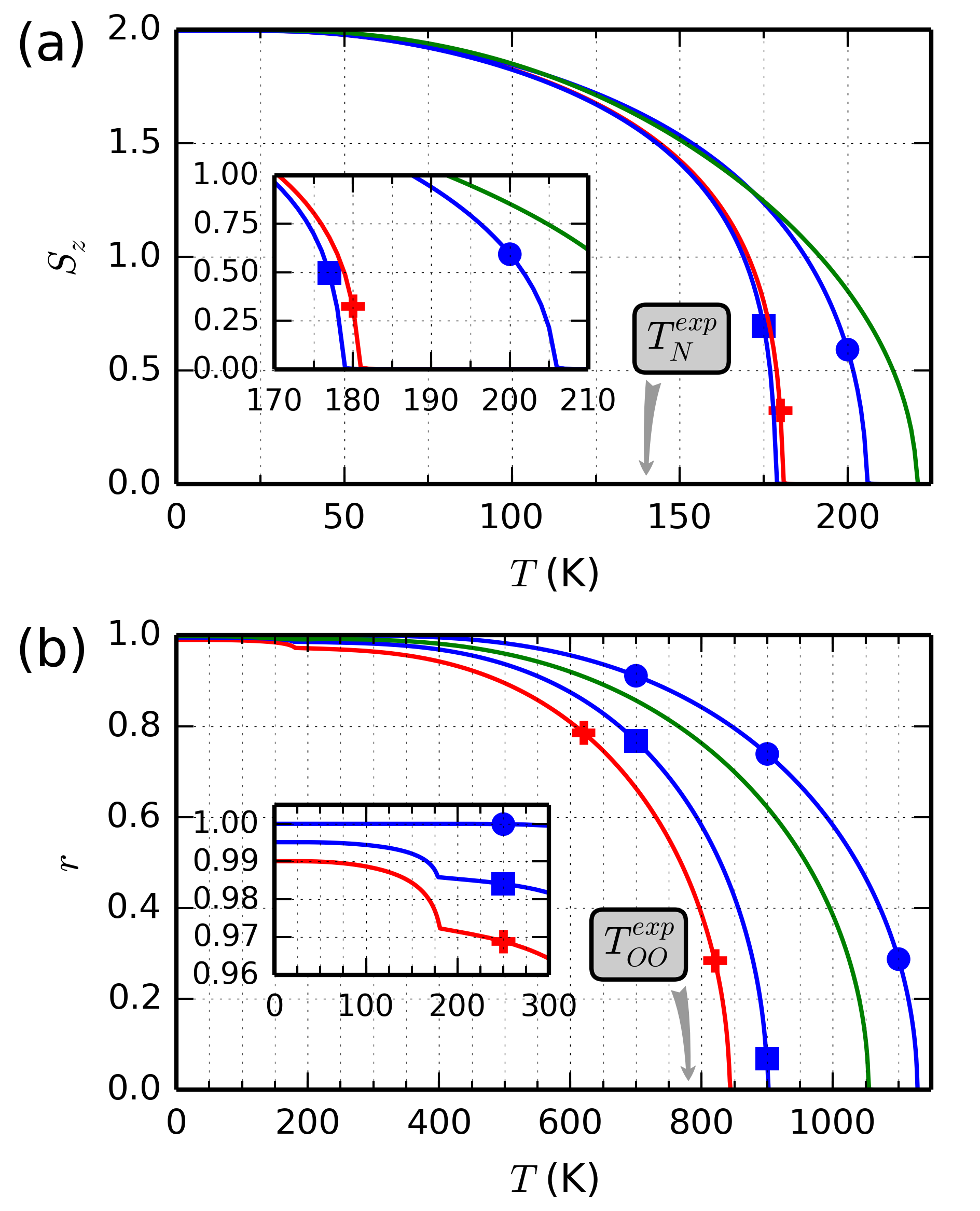}
 \caption{
Temperature dependence of the on-site parameters obtained in the
diverse calculation schemes:
(a) spin order parameter $S^z$ and
(b) orbital order parameter $r$.
The blue and red lines with markers correspond to decoupled schemes.
The temperature dependence as obtained in implementation with on-site
orbital calculations (blue, bullet makers $\bullet$), the cubic
cluster orbital calculations (blue, square makers $\blacksquare$), and
the hedgehog cluster orbital calculations (red, $+$ makers).
The green line without markers corresponds to multi-single-bond scheme.
As a reference, the experimental values of phase transition
temperatures are indicated in (a) and (b).
Parameters as in Eqs. (\ref{para}) and (\ref{kappa}).}
 \label{fig:on-site}
\end{figure}

The obtained temperature dependence of the spin order parameters is
shown in Fig. \ref{fig:on-site}(a), while the orbital order is studied
in Fig. \ref{fig:on-site}(b). In both cases
one finds a phase transition to the ordered state at temperatures
somewhat higher than the experimental values. Nevertheless, the results
are very encouraging as the MF values are significantly reduced, both
for spin and for orbital transition.

Based on the results displayed in Fig. \ref{fig:on-site}(b) we may put
forward some important statements connected with orbital-only clusters
comparison. First, we notice that there is no quantum fluctuation
within on-site calculations and that the quantum fluctuation magnitude
(at $T=0$) within the hedgehog cluster is equal to doubled the quantum
fluctuation magnitude (at $T=0$) in the cubic cluster. We suggest that
it is due to the fact that the distinguished site in the hedgehog
cluster has 6 quantum interactions, whereas in the cubic cluster all
the sites have only 3 quantum interactions. From this point of view the
hedgehog cluster is more realistic than the cubic one. Second, we may
clearly see that phase transition temperatures fulfill the relation:
$T_{\rm OO}(\text{hedgehog})<
T_{\rm OO}(\text{cubic})\ll T_{\rm OO}^{\text{MF}}$.
(The second inequality follows from including quantum fluctuations
beyond MF.) The first one provides a second argument which favors the
hedgehog cluster as treating the orbital quantum fluctuations in a more
realistic way.

At $T=0$ the orbital angle $\theta$ (\ref{theta}) is close to
$84$\textdegree\ and increases towards $T_{\rm N}$. Above the magnetic
transition for $T>T_{\rm N}$ it is almost constant and close to
$90$\textdegree\, except for the hedgehog cluster where it increases
further and tends to $180$\textdegree\ when $T\to T_{\rm OO}$
(not shown). We remark that this remarkable temperature dependence
results however in better predictions of the orbital bond correlations
characterized by
$\langle\mathcal{P}^{\zeta\zeta}_{\langle ij\rangle}\rangle$
(see below). The higher order JT terms would increase
this angle further as reported before \cite{Oka02}, similar to the
Goodenough terms in the superexchange \cite{Ole05}. Both interactions
are neglected here as we are interested in the impact of spin-orbital
entanglement on the magnetic phase transition rather than in
quantitative analysis of the experimental data.

\subsection{Spin and orbital phase transitions}
\label{sec:phd}

For assumed physical parameters values we obtain
$T^{\text{CMF}}_{\rm N}\approx 180$ K (181 K for the decoupled scheme
with the hedgehog cluster). As expected, this value is higher than the
experimental one ($T_{\rm N}^{\text{exp}}=140K$). The origin of the
difference is in cluster MF calculation scheme. To evaluate how much
the transition temperature is overestimated in MF calculation
(using our sputnik cluster),
we performed model cluster calculations for isotropic three-dimensional
(3D) Heisenberg antiferromagnet for $S=2$ with unit magnetic exchange.
One obtains the transition temperatures equal to 11.3, whereas the true
value is equal to 8.5. One may deduce the latter value using the
semiempirical formula \cite{Fle04} with pretty good accuracy.
This means that the true transition temperature value is equal to
approximately 75\% of the calculated transition temperature value. In
our case, we may correct the above $T^{\text{MF}}_{\rm N}\approx 180$
K using the same factor $0.75$ to obtain the corrected empirical value
$T^{\text{\rm emp}}_{\rm N}\approx 135$ K.
This is indeed a very good agreement with experiment.

\begin{table}[t!]
\caption{Orbital transition temperatures $T_{\rm OO}$ obtained in
various calculation schemes.
Parameters as in Eqs. (\ref{para}) and (\ref{kappa}).}
\begin{ruledtabular}
 \begin{tabular}{ccc}
 calculation scheme &   cluster   & $T_{\rm OO}$ (K) \\
  \hline
      mean field          & single site &     1128   \\
spin-orbital disentangled &  hedgehog   & $\,\, 843$ \\
spin-orbital disentangled &    cubic    & $\,\, 901$ \\
spin-orbital    entangled & single bond &     1055   \\
 \end{tabular}
\end{ruledtabular}
 \label{tab:TOO_methods}
\end{table}

The case of $T_{\rm OO}$ is more subtle. First we notice considerable
discrepancy between the predicted values of $T_{\rm OO}$ within the
various calculation schemes used, see Table \ref{tab:TOO_methods}. We
adopt here the lowest value $T^{\rm CMF}_{\rm OO}=843$ K as obtained
with hedgehog cluster. Note that this value is radically reduced by
25\% from $T^{\rm MF}_{\rm OO}=1128$ K. As in spin case, one expects
that the experimental value would be still lower but we do not have a
simple method to reduce the above value to simulate the effect of
quantum fluctuations in the orbital model. \textit{De facto}, quantum
fluctuations are much reduced in a 3D orbital system compared with the
Heisenberg spin model \cite{vdB99}, so we suggest that the correction
of the estimated $T^{\rm MF}_{\rm OO}$ would be less than 10\% which
brings it very close to experiment. This also agrees qualitatively
with the expected reduction of the MF result that would be for the 3D
orbital model significantly lower than 37\% found for $S=1/2$ spin
Heisenberg model \cite{Fle04} (which gives the corrected empirical
value $T_{\rm OO}^{\rm emp}=711$ K).

Secondly we investigate the $\kappa$ dependence of $T_{\rm OO}$
(for disentangled scheme with a hedgehog cluster).
We have found that $T_{\rm OO}\simeq 843$ K for $\kappa=6$ meV and
$T_{\rm OO}\simeq 943$ K for $\kappa=8$ meV. This implies that in the
physically interesting regime,
\begin{equation}
\frac{\partial T_{\rm OO}}{\partial\kappa}\simeq 50\,
\frac{\rm K}{\rm meV}\,.
\end{equation}

\subsection{Spin exchange constants and
$\big\langle\vec S_i\cdot\vec S_j \big\rangle$ correlations}
\label{sec:spin}

\begin{figure}[t!]
 \centering
 \includegraphics[width=\columnwidth]{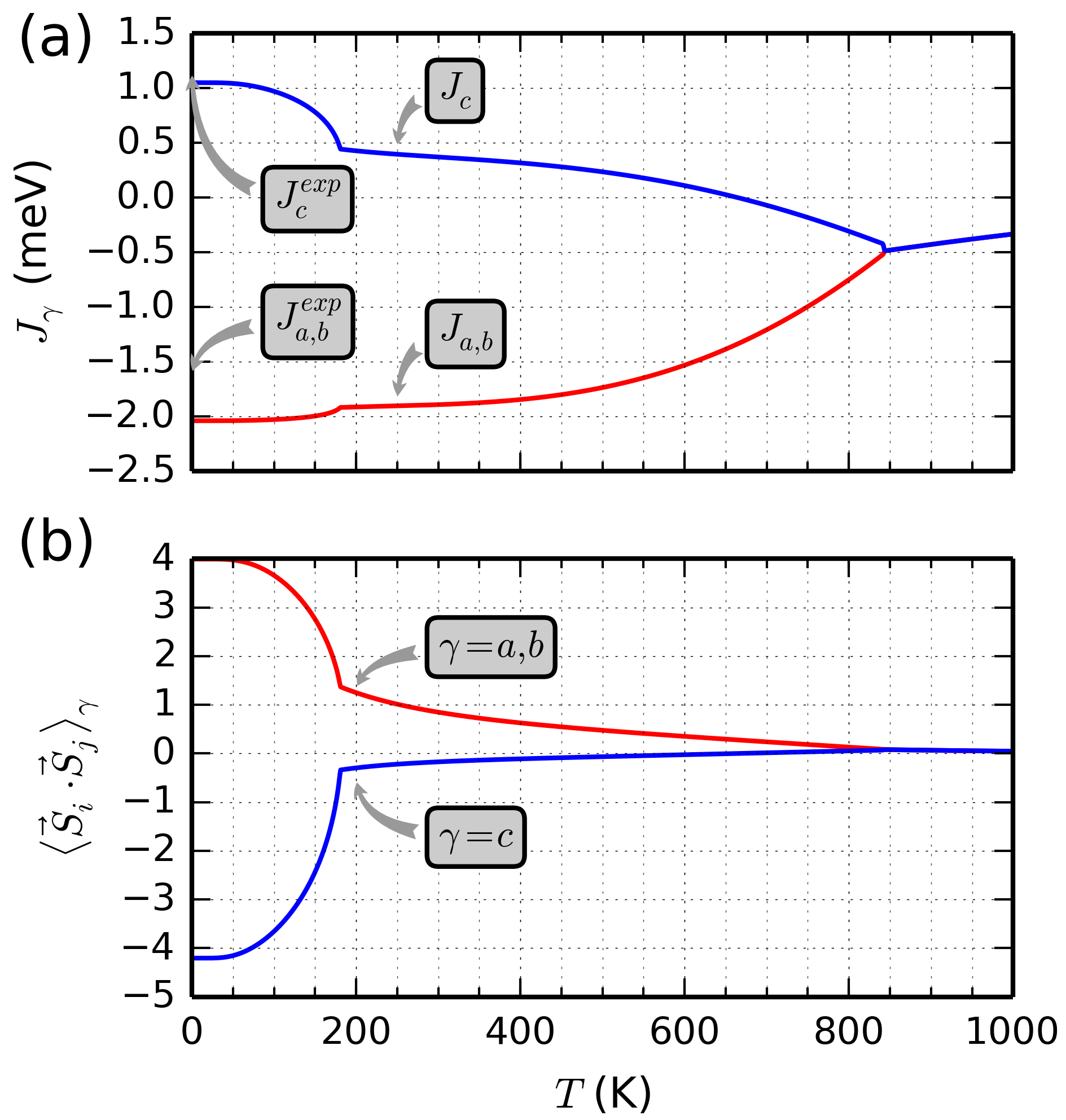}
\caption{Temperature dependence of the effective spin model obtained
from the decoupled (spin-orbital disentangled) model
(in the hedgehog orbital cluster implementation):
(a) spin exchange constants $J_{ab}$ and $J_c$ and
(b) spin-spin correlations $\langle\vec S_i\cdot\vec S_j\rangle$ on
nearest neighbor bonds $\langle ij\rangle$ within the $(a,b)$ planes
and along the $c$ axis.
Parameters as in Eqs. (\ref{para}) and (\ref{kappa}).}
 \label{fig:DecoupledParams}
\end{figure}

In the \textit{decoupling scheme} the values of $J_\gamma$'s constants
are computed in a self-consistent way. The temperature dependence of
the outcome exchange constants $J_{ab}$ and $J_c$ is shown in Fig.
\ref{fig:DecoupledParams}(a). As expected for the $A$-AF order, one
finds $J_{ab}<0$ and $J_c>0$. The agreement with the experimental
values is fair and we conclude that the overall description of the
magnetic interactions in LaMnO$_3$ is consistent with experiment. We
would like to emphasize that unlike in pure spin systems, here the
exchange constants and tuned by orbital correlations and thus depend
on temperature, both below and above $T_{\rm N}$. When $C$-AO exists,
they are anisotropic but only below $T_{\rm N}$ the positive value of
$J_c$ is boosted when the orbitals change below the magnetic transition.
Above $T_{\rm OO}$ the spin exchange interactions are isotropic and
weakly negative.

In a similar way the self-consistent values of spin-spin correlations
for nearest neighbors, $\langle\vec S_i\cdot\vec S_j\rangle$, are
obtained for $\langle ij\rangle\parallel\gamma$, for $\gamma=a,b$ and
$\gamma=c$. Figure \ref{fig:DecoupledParams}(b) shows the corresponding
temperature dependence. Note that for $T=0$ for the bonds along the $c$
axis (the one with AF coupling $J_c>0$)
$\langle\vec S_i\cdot\vec S_j\rangle\simeq -4.21$ which indicates some
but not large quantum fluctuations. The above value belongs to the
range between classical N\'eel value $-4.0$ and quantum singlet value
for $S=2$ spins, $-6.0$. It is not surprising that due to imposed
symmetry breaking the obtained value of $-4.21$ is much closer to the
classical N\'eel state, although the quantum fluctuations are still
considerable.

\subsection{On-bond orbital correlations}
\label{sec:orbi}

Above $T_{rm N}$ we find $C$-AO order up to $T^{\rm CMF}_{\rm OO}$.
The orbital model for $e_g$ orbitals is more classical than the $S=1/2$
Heisenberg model \cite{vdB99} but quantum effects are also important
\cite{Ryn10}. Similar to the two-dimensional compass model where the
pseudospins are entangled \cite{Cza16}, one expects that the quantum
effects for $e_g$ orbital model contribute both to the intersite
orbital correlations and to the phase transition temperature
$T_{\rm OO}$. This motivates investigation of
$\mathcal{P}_{\langle ij\rangle}$'s operators (\ref{zetxi})-(\ref{xixi})
and comparing their actual mean values with the corresponding values
given by the product of MF values of $P$'s operators at sites $i$ and
$j$. This study provides information about the on-bonds orbital
entanglement. Here, we alias the former as true mean value and the
latter as slave MF. For example, for
$\mathcal{P}^{\zeta\xi}_{\langle ij \rangle}$ operator Eq. (\ref{zetxi})
one has to compare:
\begin{displaymath}
 \begin{array}{cl}
\text{mean value} &\langle\mathcal{P}^{\zeta\xi}_{\langle ij \rangle}\rangle, \\
\text{MF \textit{Ansatz}}&
2\langle P^{\zeta}_{(i}\rangle\langle P^{\xi}_{j)}\rangle.
 \end{array}
\end{displaymath}
We use decoupled calculation scheme with the hedgehog orbital cluster
implementation to obtain these mean values.

\begin{figure}[t!]
 \centering
 \includegraphics[width=\columnwidth]{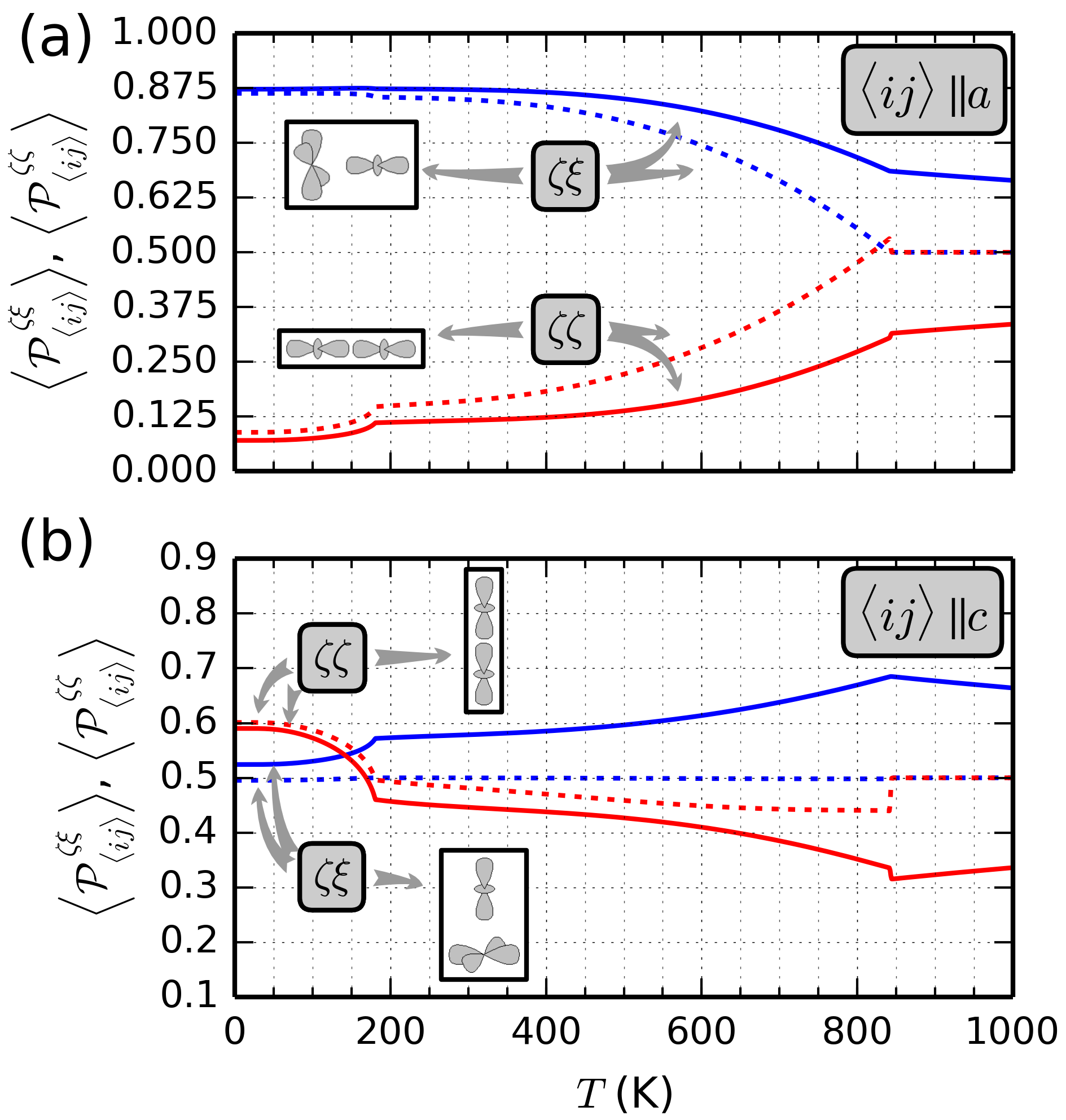}
\caption{The true mean values:
$\big\langle\mathcal{P}^{\zeta\xi}_{\langle ij\rangle}\big\rangle$ and
$\big\langle\mathcal{P}^{\zeta\zeta}_{\langle ij\rangle}\big\rangle$
(solid lines) and slaves MF values:
$2\big\langle P^\zeta_{(i}\big\rangle\langle P^\xi_{j)}\big\rangle$ and
$2\big\langle P^\zeta_{i}\big\rangle\langle P^\zeta_{j}\big\rangle$
(dashed lines).
All results are from decoupled (disentangled) calculation scheme
with the hedgehog orbital cluster implementation.
Parameters as in Eqs. (\ref{para}) and (\ref{kappa}).}
 \label{fig:mathcalP}
\end{figure}

The thermal dependence for both, the true and the factorized slave MF
values of on-bound orbital projection operators,
$\langle\mathcal{P}^{\zeta\zeta}_{\langle ij\rangle}\rangle$ and
$\langle\mathcal{P}^{\zeta\xi}_{\langle ij\rangle}\rangle$,
are shown in Fig. \ref{fig:mathcalP}. Note, that the third on-bond
orbital projection operator,
$\langle\mathcal{P}^{\xi\xi}_{\langle ij\rangle}\rangle$,
does not contribute to $H_J$ (\ref{HJ}) and is therefore not shown.

In case of bonds parallel to the $a$ axis the most important factor
that governs the $\mathcal{P}$'s operators mean values is classical
correlation that follows directly from the on-site orbital pattern
(with $\theta\approx90$\textdegree). One finds that below $T_{\rm N}$
the $\mathcal{P}$'s values are close to their classical MF values
($r=1$) with $\theta=90$\textdegree.
(The corresponding classical values are equal to $0.875$ and $0.125$.)
In range from $T_{\rm N}$ to $T_{\rm OO}$ the parameter $r$ decreases
monotonically down to $0$. The slave $\mathcal{P}$'s MF values behavior
mimics this temperature dependence of the $r$ parameter. Hence slave
$\mathcal{P}$'s MF values approach the high temperature limit
$\frac{1}{2}$ while temperature increases $T\to T_{\rm OO}$ and take
this limiting value when the orbitals are completely disordered for
$T\to T_{\rm OO}$. In contrast, short-range correlations which may be
seen as on-bond entanglement are enhanced near the orbital phase
transition and the values of both bond projection operators,
$\mathcal{P}^{\zeta\zeta}_{\langle ij\rangle\parallel ab}$ and
$\mathcal{P}^{\zeta\xi}_{\langle ij\rangle\parallel ab}$ are
significantly smaller/larger than $\frac{1}{2}$ for
$T\approx T_{\rm OO}$. At $T_{\rm OO}$ the difference between true mean
value and $\frac{1}{2}$ found in MF is approximately equal to $0.15$.

In case of bonds parallel to the $c$ axis the situation is different.
Within assumed orbital pattern with $\theta\approx\pm90$\textdegree\
the directional $|3z^2-r^2\rangle/\sqrt{6}$ state is just as likely as
the planar $|x^2-y^2\rangle/\sqrt{2}$ state. This is the reason why the
MF slave mean values of $\mathcal{P}$'s operators are here quite close
to high temperature limit value $\frac{1}{2}$ (even at $T=0$). Once
more, the on-bond entanglement makes the true mean value more distant
from $1/2$ in comparison to slave ones. Despite that, the true mean
values are quite close but deviate from $1/2$.
At $T_{\rm OO}$ the difference between true mean value and
$1/2$ is as big as approximately $0.20$.
On-bond entanglement boosts the contribution of the configuration
$|z^2\rangle|z^2\rangle$ for sites on-bond.
This is reasonable, as there is an interaction
$\propto\mathcal{P}^{\zeta\zeta}_{\langle ij\rangle\parallel c}$ which
stabilizes this orbital configuration.

\subsection{Spectral weights}
\label{sec:sw}

A detailed experimental study of the optical spectral weights for HS
and LS states was presented in Ref. \cite{Kov10}. The experimental data
were obtained up to room temperature and show gradual reduction of the
weight in the HS channel and an accompanying increase of the LS part
for the $b$ axis polarization. For a Mott insulator the optical
spectral weights can be determined from the superexchange terms
$H_{J,n}^{(\gamma)}$ which stem from charge excitation $n$ to either
HS or LS states along a bond in the direction $\gamma$ \cite{Kha04}.
Such terms are related to the optical spectral weights via the
kinetic energy,
\begin{equation}
K_n^{(\gamma)}=-2\left\langle H_{J,n}^{(\gamma)}\right\rangle.
\end{equation}
These contributions depend both on the cubic direction $\gamma$ and on
the type of charge transfer excitation $n$. The dimensionless optical
spectral weights for HS and LS part are proportional to
$K_n^{(\gamma)}$,
\begin{eqnarray}
N_{eff}(\rm HS)&=&\frac{ma_0^2}{\hbar^2}\,K_1^{(\gamma)},\\
N_{eff}(\rm LS)&=&\frac{ma_0^2}{\hbar^2}\,\sum_{n>1}K_n^{(\gamma)}.
\end{eqnarray}
We used the inverse volume $a_0^{-3}=1.7\cdot10^{22}$ cm$^{-3}$
\cite{Kov10}.

\begin{figure}[t!]
 \centering
 \includegraphics[width=\columnwidth]{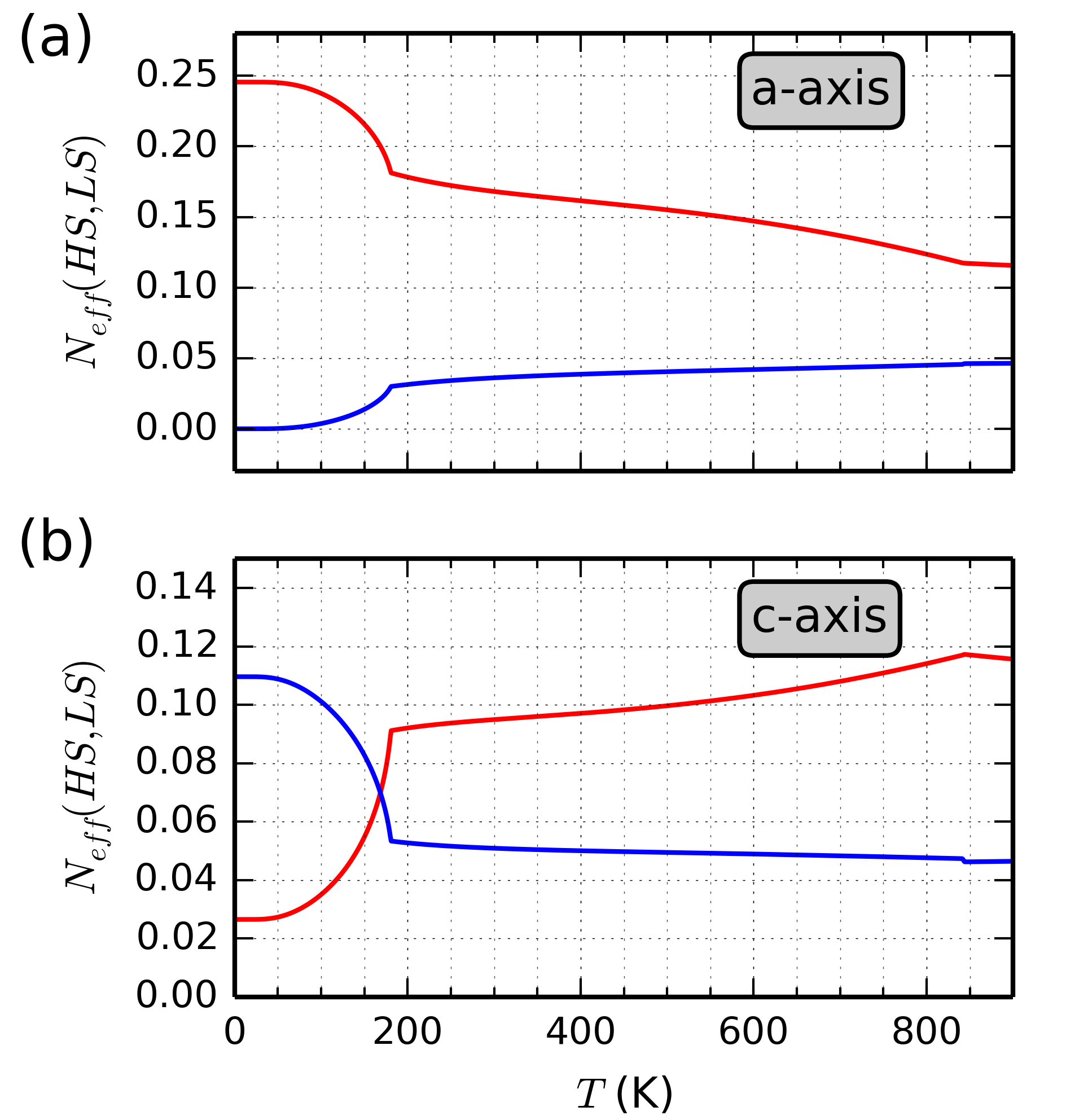}
\caption{
Optical spectral weights $N_{eff}(\rm HS)$ (red lines) and
$N_{eff}(\rm LS)$ (blue lines) obtained for the polarization along the:
(a) $a$ axis and
(b) $c$ axis.
All results are from decoupled (disentangled) calculation scheme
with the hedgehog orbital cluster implementation.
Parameters as in Eqs. (\ref{para}) and (\ref{kappa}).}
 \label{fig:sw}
\end{figure}

We present below the numerical results obtained in the entire
temperature range up to the orbital transition at $T_{\rm OO}$, see
Fig. \ref{fig:sw}. It is not surprising that the HS spectral weight
dominates for the polarization along the $a$ axis at low temperature
when $A$-AF order persists. Its decrease from $T=0$ to $T=T_{\rm N}$
is only by about 25\%, and next the weight decreases steadily towards
the orbital transition at $T_{\rm OO}$, see Fig. \ref{fig:sw}(a). This
variation of the HS part is accompanied by gradual increase of the LS
part in the entire temperature range.

At low temperature the role of HS and LS spectral weights is reversed
for the AF bonds along the $c$ axis, see Fig. \ref{fig:sw}(b). But in
contrast to the FM bonds shown in Fig. \ref{fig:sw}(a), the changes
when $T_{\rm N}$ is approached from $T=0$ are here more pronounced and
the LS spectral weight drops from $\sim 0.11$ at $T=0$ to $\sim 0.05$
at $T_{\rm N}$, while simultaneously the HS spectral weight increases
from $\sim 0.03$ to $\sim 0.09$, i.e., by a factor close to 3.
It is quite remarkable that the spectral weight for the HS part is
here much higher than the that of the LS part in spite of the onset
of the AF order at low temperature. We also note that the LS spectral
weight practically does not change when temperature increases within
the interval $T\in[T_{\rm N},T_{\rm OO}]$, and the LS weight is almost
isotropic above $T_{\rm OO}$. Most importantly, we observe that:
(i) the HS part dominates and is close to 0.11, while
(ii) the LS part is much lower and close to 0.05
for both $a$ and $c$ bond polarization above $T_{\rm OO}$ [note
different vertical scales in Figs. \ref{fig:sw}(a) and \ref{fig:sw}(b)].
Thus, the distribution of optical spectral weights becomes isotropic
above $T_{\rm OO}$ and the spectral weight of the HS part is more
than twice larger than that of the LS part.

\section{SPIN-ORBITAL ENTANGLEMENT}
\label{sec:enta}

\subsection{On-site spin-orbital entanglement}
\label{sec:site}

\begin{figure}[t!]
 \centering
 \includegraphics[width=8.2cm]{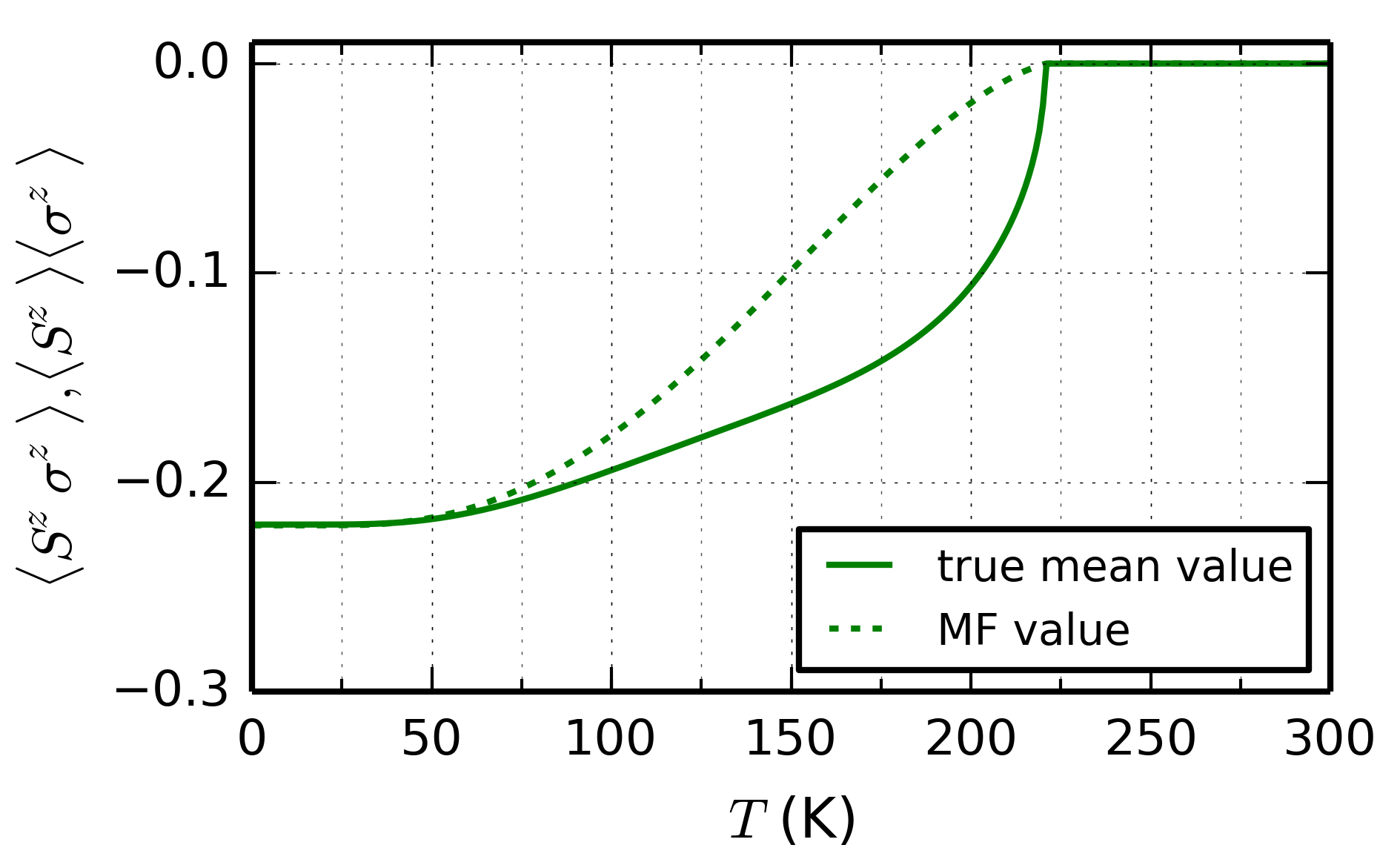}
\caption{The calculated on-site mean value
$\langle S^z_i\sigma^z_i\rangle$ (solid lines) and its MF estimated
value $\langle S^z_i\rangle\langle\sigma^z_i\rangle$ (dashed lines).
All results are from multi-single-bond calculation scheme, see Sec.
\ref{sec:bond}. Parameters as in Eqs. (\ref{para}) and (\ref{kappa}).}
 \label{fig:on_site}
\end{figure}

A fundamental problem for spin-orbital systems is weather spin and
orbital operators can be disentangled, as we have implemented for all
the calculations presented so far but the calculations of a single
bond. Below we analyze the on-site spin-orbital entanglement in a way
similar to that used for the orbital entanglement in the previous
Section. We compare the calculated spin-orbital correlation,
$\langle S^z_i\sigma^z_i\rangle$ with its factorized MF value,
$\langle S^z_i\rangle\langle\sigma^z_i\rangle$. The results are shown
in Fig. \ref{fig:on_site}. Both on-site
quantities are finite only below the magnetic transition at $T_{\rm N}$
and they also agree for the ground state at $T=0$.

However, one finds that some discrepancy between the actual value of
$\langle S^z_i\sigma^z_i\rangle$ and its MF estimate develops in the
temperature interval from approximately half of $T_{\rm N}$ to
$T_{\rm N}$. This may result in some amplification of the value of
$T_{\rm N}$ itself over the MF value although we should point out that
the considered on-site $\langle S^z_i\sigma^z_i\rangle$ mean value may
take any value from the interval $[-2.0,2.0]$ whereas it is here close
to $-0.22$ at $T=0$, so we may consider it as being rather small.
Furthermore, one finds that the factorized average drops faster towards
zero and is already rather close to it for $T>150$ K whereas the direct
calculation gives a rather different behavior, with a rapid drop of
$\langle S^z_i\sigma^z_i\rangle$ close to $T_{\rm N}\simeq 220$ K.

We have also performed similar calculations for the average
$|\langle S^z_i\sigma^x_i\rangle|$. Its value is much larger and close
to $\sim 2.0$ at $T=0$ but one finds no discrepancy between the direct
calculation and the approximate result obtained by the MF factorization
here and also close to $T_{\rm N}$. Altogether we conclude that on-site
spin-orbital entanglement is of little importance.

\subsection{On-bond spin-orbital entanglement}
\label{sec:bond}

\begin{figure}[t!]
 \centering
\includegraphics[width=\columnwidth]{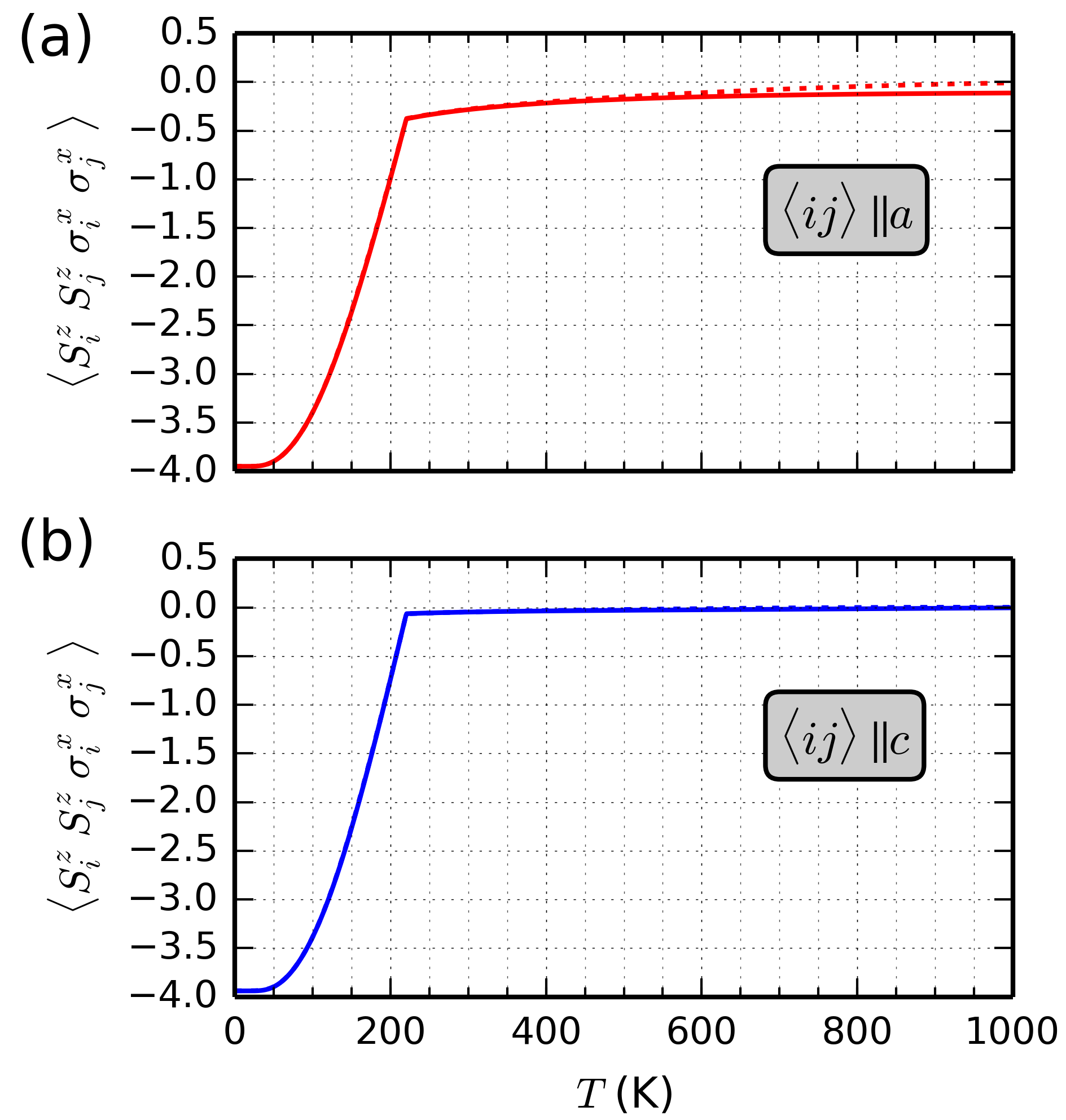}
\caption{
The calculated spin-orbital averages
$\langle S^z_iS^z_j\sigma^x_i\sigma^x_j\rangle$ (solid lines)
and the corresponding disentangled MF values,
$\langle S^z_iS^z_j\rangle\langle\sigma^x_i\sigma^x_j\rangle$ (dashed
lines) as obtained from multi-single-bond calculation scheme for a
bond:
(a) within the $(a,b)$ plane and
(b) along the $c$ axis.
Parameters as in Eqs. (\ref{para}) and (\ref{kappa}).}
 \label{fig:entx}
\end{figure}

The occupied $e_g$ orbitals are characterized by the angle
$\theta\approx 45$\textdegree, i.e., the occupied states are rather
close to the eigenstates of $\sigma^x$ operator. Therefore, assuming
that the symmetry broken states have $\langle S_i^z\rangle\neq 0$,
the relevant spin-orbital correlation function,
$\langle S^z_iS^z_j\sigma^x_i\sigma^x_j\rangle$, is expected to be
large and negative for all the bonds. The negative value is obtained
when either spins are AF and orbitals are the same for the bonds
$\langle ij\rangle\parallel c$, or spins are FM but orbitals alternate
in the $(a,b)$ planes for $\langle ij\rangle\parallel a$. Indeed, one
finds that these mixed on-bond correlations are close to $-4.0$
independently of the bond direction, see Fig. \ref{fig:entx}. These
correlation functions are almost equal to the products of spin and
orbital correlations, i.e.,
\begin{equation}
\langle S^z_iS^z_j\sigma^x_i\sigma^x_j\rangle\simeq
\langle S^z_iS^z_j\rangle\langle\sigma^x_i\sigma^x_j\rangle.
\end{equation}
Therefore, spin and orbital correlations are almost disentangled.

\begin{figure}[t!]
 \centering
\includegraphics[width=\columnwidth]{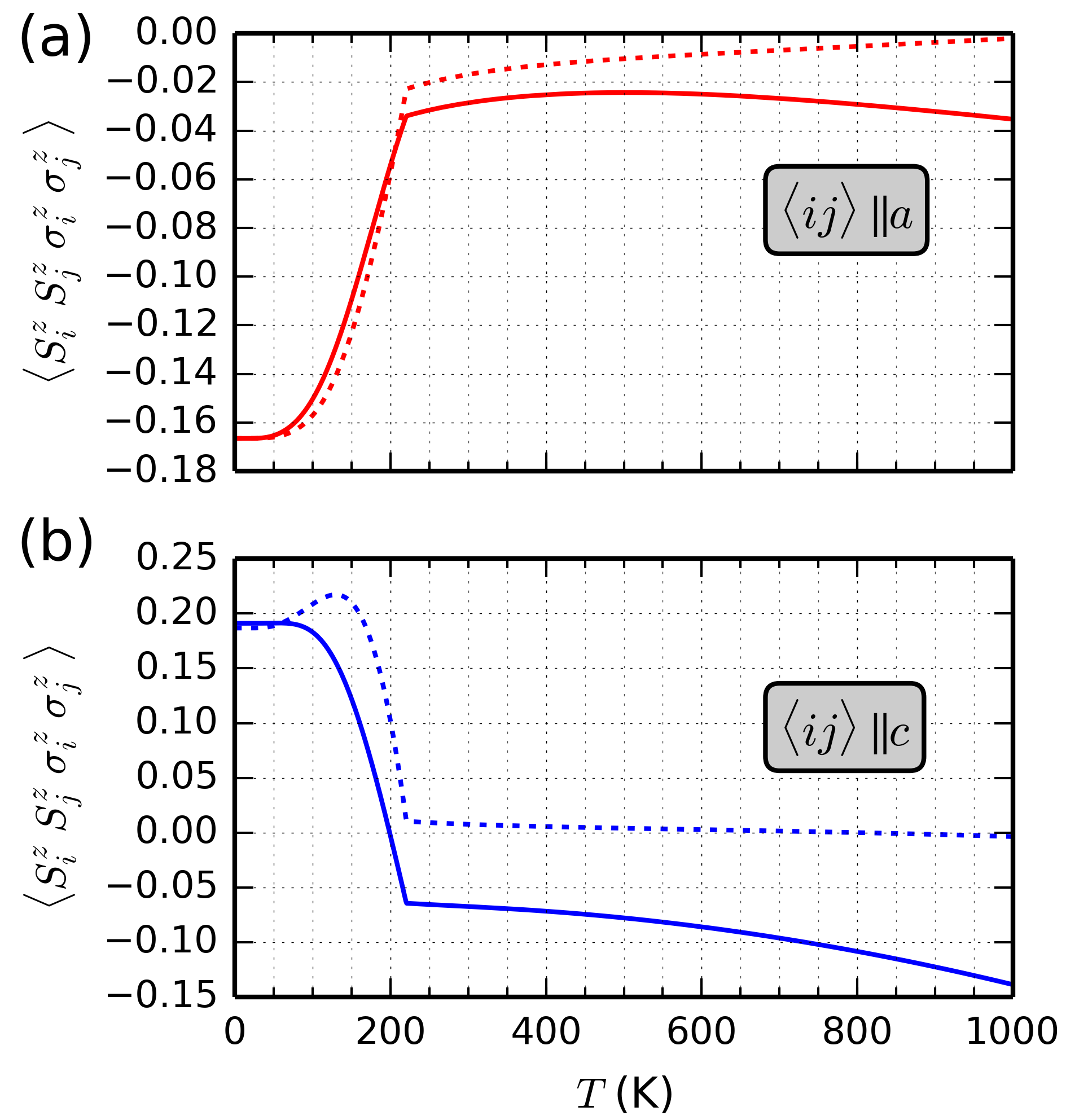}
\caption{
The calculated spin-orbital averages
$\langle S^z_iS^z_j\sigma^z_i\sigma^z_j\rangle$ (solid lines)
and the corresponding disentangled MF values,
$\langle S^z_iS^z_j\rangle\langle\sigma^z_i\sigma^z_j\rangle$ (dashed
lines) as obtained from multi-single-bond calculation scheme for a
bond:
(a) within the $(a,b)$ plane and
(b) along  the $c$     axis.
Parameters as in Eqs. (\ref{para}) and (\ref{kappa}).}
 \label{fig:entz}
\end{figure}

The on-bond spin-orbital correlations which involve $\sigma_i^z$
operators, $|\langle S^z_iS^z_j\sigma^z_i\sigma^z_j\rangle|$,
are much smaller in the entire temperature range, see Fig.
\ref{fig:entz}, than those involving $\sigma_i^x$, shown in Fig.
\ref{fig:entx}. At $T=0$ these averages are equal to the products
of spin and orbital terms, i.e., they are also disentangled.
In contrast, the result obtained for $T>T_{\rm N}$ is qualitatively
new. In this regime the orbital correlations
$\langle\sigma^z_i\sigma^z_j\rangle$ are close to zero which explains
why the MF values are almost zero for both directions,
$\langle ij\rangle\parallel a$ and $\langle ij\rangle\parallel c$,
see Figs. \ref{fig:entz}(a) and \ref{fig:entz}(b).
This is contrast to large entanglement for $S=1/2$ spins which may
trigger new phases either in the limit vanishing Hund's exchange
\cite{You15} or when spin interactions change the sign in the
Kugel-Khomskii model \cite{Brz12}.

However, we observe that some weak spin-orbital fluctuations occur.
The spin-spin correlation functions $\langle S^z_iS^z_j\rangle$, see
Fig. \ref{fig:DecoupledParams}(b), are much smaller than within the
$A$-AF phase but are finite and follow the sign of the respective
exchange constant $J_{ab}$ and $J_c$ above $T_{\rm N}$ --- they reflect
short-range spin order. Orbital order is robust and persists above
$T_{\rm N}$, with $\langle\sigma^x_i\sigma^x_j\rangle\simeq\pm 1.0$ and
$\langle\sigma^z_i\sigma^z_j\rangle\simeq\pm 0$, but weak spin-orbital
fluctuations occur there and thus one finds for both bond directions
shown in Fig. \ref{fig:entz} that
$\langle S^z_iS^z_j\sigma^z_i\sigma^z_j\rangle<0$.
These finite values result from the superposition of spin short-range
order with orbital thermal fluctuations. As a result the difference
between the spin-orbital on-bond correlation functions and the
respective products of spin and orbital MF values steadily increase
above $T_{\rm N}$. Altogether we conclude that spin-orbital
correlations are almost completely disentangled in the ground state
but joint spin-orbital fluctuations are activated by temperature.

\subsection{The impact of entanglement on N\'eel temperature}
\label{sec:TN}

The results above vividly show that, in case of finite temperature,
there are some nontrivial correlations between:
(i) a spin state and an orbital state at each site, and similarly
(ii) orbital states of adjacent sites.
Although we quantitatively describe the difference between the values
of  pertinent expressions of $\langle AB\rangle$ type and the
corresponding expressions of $\langle A\rangle\langle B\rangle$ type,
the results give us little information about the influence
of corresponding correlations on the predicted magnetic transition
temperature $T_{\rm N}$. As the result it is difficult to rate their
physical importance.

To examine the influence of the correlations we have carried out the
following numerical experiment. We involved the model in which the
correlations are taken into account and calculated the value of
$T_{\rm N}$ at this level of the theory. Then we modified the model in
such a way that all pertinent expressions of $\langle AB\rangle$ type
were changed into decoupled expressions,
$\langle A\rangle\langle B\rangle$. Finally, we calculated the N\'eel
temperature $T_{\rm N}$ in the latter model. As the only difference
between the two results steams from the deliberately neglected
correlation, we perceive the outgoing temperature difference as
heuristic correction of $T_{\rm N}$ associated with the correlation.
The relative value of the correction gives rise to a quantitative
rate of the physical importance of the correlation.

For case (i) we have involved simple on-site model in which
\textit{both} spin and orbital degrees of freedom were introduced
collectively (thus dimension of the Hilbert space is equal to 10 as it
is a product of a spin quintet and an orbital doublet). Note that the
model is different from the on-site model described earlier in the
text (that works in \textit{decoupled calculations scheme}. The site
was immersed in the mean field of its surroundings (constructed in
line with the $A$-AF and $C$-AO pattern). In the model values
$\langle S_z\sigma_z\rangle$ occur as a part of description of the
site's surroundings. One can replace $\langle S_z\sigma_z\rangle$ by
$\langle S_z\rangle\langle\sigma_z\rangle$ with ease.

For case (ii) we have involved the "decoupled calculations" with the
hedgehog orbital cluster. In this calculation values
$\langle P^{\zeta_\gamma}_i P^{\zeta_\gamma}_j\rangle$ (and analogous)
occur as a part of the formula for magnetic exchange constants
$J_\gamma$, see Eq. (\ref{J}). Once more, one can easily replace
$\langle P^{\zeta_\gamma}_i P^{\zeta_\gamma}_j\rangle$
by the factorized product,
$\langle P^{\zeta_\gamma}_i\rangle\langle P^{\zeta_\gamma}_j\rangle$.

First, for case (i) we obtained the following $T_{\rm N}$ values:
238 K for the full model and 218K for the model with the artificially
broken on-site spin-orbital correlations. Of course both results are
overestimated as the model consists of only one site. Nevertheless,
the difference should not suffer from this problem to large extent due
to the errors' cancelation. Quite unexpectedly, the latter model has
lower transition temperatures. We can elucidate this on the basis of
the result presented in Fig. \ref{fig:on_site}. The absolute value of
$\langle S_z\rangle\langle\sigma_z\rangle$ is lower than the absolute
value of $\langle S_z\sigma_z\rangle$ and hence the interaction
between the site and its surroundings is weaker. It turns out, that
the corresponding difference in $T_N$ is as big as 20K, about 10\%
of the total value.
Second, for case (ii) we obtained the following $T_{\rm N}$ values:
181K for the full model and 201K for model with the artificially
broken on-bound orbital correlations. The difference is equal to 20K.

We may conclude, that the both described correlations have significant
impact on N\'eel temperature (close to 10\% of the actual $T_{\rm N}$
value). However, corresponding corrections have opposite sign. On the
grounds of this result we may draw the conclusion that the simplistic
disentangled (decoupled) on-site model discussed in earlier chapters
works surprisingly well as the two corrections approximately cancel out.

\section{Summary and conclusions}
\label{sec:summa}

We have investigated spin-orbital order in stoichiometric LaMnO$_3$
using the model reported earlier \cite{Fei99} which consists of
superexchange and Jahn-Teller induced orbital terms. As a preliminary,
the simple argumentation was presented that leads to quite detailed
(low temperature) electronic state description. It was elucidated that
the model used accounts for observed experimentally coexisting $A$-AF
and $C$-AO order below the N\'eel temperature, and reproduces phases
observed experimentally to some extent. A certain discrepancy between
the predicted orbital state and the order deduced from the experimental
data was detected in the value of orbital mixing angle $\theta$.
Some key suggestions were given how to enrich the model --- these
suggestions include crystal-field effect and anharmonicity of crystal
vibrations. We suggest that these extensions of the present model
should be taken into account in future studies.
Despite the elucidated discrepancy, for the time being, we suggest the
presented analysis of electronic states within this model as
insightful and instructive.

To achieve a more realistic description of the LaMnO$_3$ electronic
state, we introduced some clusters (entities taken out of the crystal)
and performed unbiased cluster mean field calculations at finite
temperature. This makes, to the best of our knowledge, the first
attempt to treat the spin and orbital phase transitions simultaneously
in such a complex perovskite system. Cluster approach allows to
evaluate the magnetic and the orbital transition temperatures in a
more realistic way than within the simple on-site mean field approach.
The obtained N\'eel temperature $T_{\rm N}\simeq 181$ K is in good
agreement with experiment, taking into account the mean field nature
of the cluster method.

Based on the cluster calculations the value of the effective
Jahn-Teller orbital coupling constant was deduced as reproducing
correctly the orbital transition temperature. Taking realistic values
for all the other parameters, we find $\kappa\simeq 6$ meV in Eq.
(\ref{HJT}). Although such orbital interactions contribute on equal
footing as those which result from superexchange processes for the
$e_g$ orbital order, they are reduced from those estimated before due
to the present cluster mean field approach.

By a careful analysis of entanglement, we have shown that both
entangled on-site spin-orbital correlations and intersite
orbital-orbital correlations influence the value of the N\'eel
temperature $T_{\rm N}$. However, when on-site spin-orbital
entanglement is included, $T_{\rm N}$ increases, while it decreases
by a similar amount when intersite orbital-orbital correlations are
not factorized. Therefore, the errors generated by decoupling such
operators partly compensate each other which explains the apparent
success of the simplified effective mean field model for LaMnO$_3$
where on-site spin-orbital and intersite orbital-orbital correlations
are neglected.

It would be interesting to perform similar analysis of entanglement
for other spin-orbital models. It is known that entanglement is large
for small $S=1/2$ spins as shown for SU(2)$\otimes$SU(2) models
\cite{Ole06} and for more general interactions in one dimension
\cite{You15}. Here a study of spin-orbital interactions in iron
pnictides \cite{Kru09} would be of interest as for $S=1$ spins one
expects larger entanglement than in LaMnO$_3$.
Experimental consequences could be quite challenging.

Finally,
we emphasize that the present realistic spin-orbital superexchange
model has to be treated beyond the on-site mean field to extract from
it realistic values of magnetic exchange constants. The obtained
values of $J_{ab}$ and $J_c$ explain the spin excitations in the
observed $A$-type antiferromagnetic phase. We have presented also the
optical spectral weights deduced from the presented model in a broad
range of temperature.

Summarizing, we have shown evidence that spin-orbital entanglement is
rather weak in LaMnO$_3$. The performed cluster mean field analysis
allows us to establish that the spin-orbital entanglement is small for
both on-site and on-bond quantities. This follows from large $S=2$
spins and explains why calculations based on the separation of spin
and orbital degrees of freedom are so successful in providing valuable
insights into the experimentally observable quantities for LaMnO$_3$.
The most important prediction of the present theory is that the
spectral weights become isotropic above the orbital transition
temperature $T_{\rm OO}$ and the high-spin processes dominate over
low-spin ones and give much higher spectral weight at low energy.
This prediction could be of importance not only for LaMnO$_3$ but
also for (LaNiO$_3)_n$/(LaMnO$_3)_2$ superlattices investigated
recently \cite{DiP15}.

\acknowledgments

We thank Louis Felix Feiner for insightful discussion.
We kindly acknowledge support by Narodowe Centrum Nauki
(NCN, National Science Center) under Project No. 2012/04/A/ST3/00331.


\begin{thebibliography}{99}

\bibitem{Tok00} Y. Tokura and N. Nagaosa,
                   Orbital physics in transition metal oxides,
                   Science \textbf{288}, 462 (2000).

\bibitem{Goode} J.-S. Zhou and J. B. Goodenough,
                   Unusual Evolution of the Magnetic Interactions versus
                   Structural Distortions in RMnO$_3$ Perovskites,
                   Phys. Rev. Lett. \textbf{96}, 247202 (2006).

\bibitem{Fei97} L. F. Feiner, A. M. Ole\'s, and J. Zaanen,
                   Quantum melting of magnetic order due to orbital fluctuations,
                   Phys. Rev. Lett. \textbf{78}, 2799 (1997);
                   Quantum disorder versus order-out-of-disorder
                   in the Kugel-Khomskii model,
                   J.~Phys.: Condens. Matter \textbf{10}, L555 (1998).

\bibitem{Kha00} G. Khaliullin and S. Maekawa,
                   Orbital liquid in three-dimensional Mott insulator: LaTiO$_3$,
                   Phys. Rev. Lett. \textbf{85}, 3950 (2000).

\bibitem{Kha01} G. Khaliullin, P. Horsch, and A. M. Ole\'s,
                   Spin order due to orbital fluctuations: Cubic vanadates,
                   Phys. Rev. Lett. \textbf{86}, 3879 (2001).

\bibitem{Kha04} G. Khaliullin, P. Horsch, and A. M. Ole\'s,
                   Theory of optical spectral weights in Mott insulators
                   with orbital degrees of freedom,
                   Phys. Rev. B \textbf{70}, 195103 (2004).

\bibitem{Kha05} G. Khaliullin,
                   Orbital order and fluctuations in Mott insulators,
                   Prog. Theor. Phys. Suppl. \textbf{160}, 155 (2005).

\bibitem{Kug82} K. I. Kugel and D. I. Khomskii,
                   The Jahn-Teller effect and magnetism:
                   Transition metal compounds,
                   Usp. Fiz. Nauk \textbf{136}, 621 (1982)
                  [Sov. Phys. Usp. \textbf{25}, 231 (1982)].

\bibitem{Ole05} A. M. Ole\'s, G. Khaliullin, P. Horsch, and L. F. Feiner,
                   Fingerprints of spin-orbital physics in Cubic Mott insulators:
                   Magnetic exchange interactions and optical spectral weights,
                   Phys. Rev. B \textbf{72}, 214431 (2005).

\bibitem{Hor08} P. Horsch, A. M. Ole\'s, L. F. Feiner, and G. Khaliullin,
                   Evolution of spin-orbital-lattice coupling in the
                   $R$VO$_3$ perovskites,
                   Phys. Rev. Lett. \textbf{100}, 167205 (2008).

\bibitem{Woh11} K. Wohlfeld, M. Daghofer, S. Nishimoto, G. Khaliullin,
                   and J. van den Brink,
                   Intrinsic Coupling of Orbital Excitations
                   to Spin Fluctuations in Mott Insulators,
                   Phys. Rev. Lett. \textbf{107}, 147201 (2011);
                P. Marra, K.~Wohlfeld, and J. van den Brink,
                   Unraveling Orbital Correlations with Magnetic
                   Resonant Inelastic X-Ray Scattering,
                   \textit{ibid.} \textbf{109}, 117401 (2012);
                V. Bisogni, K. Wohlfeld, S. Nishimoto, C. Monney,
                   J. Trinckauf, K. Zhou, R.~Kraus, K.~Koepernik,
                   C. Sekar, V.~Strocov, B. B\"uchner, T.~Schmitt,
                   J. van den Brink, and J.~Geck,
                   Orbital Control of Effective Dimensionality:
                   From Spin-Orbital Fractionalization to Confinement
                   in the Anisotropic Ladder System CaCu$_2$O$_3$,
                   \textit{ibid.} \textbf{114}, 096402 (2015);
                E. M. Plotnikova, M.~Daghofer, J.~van den Brink, and K. Wohlfeld,
                   Jahn-Teller Effect in Systems with Strong
                   On-Site Spin-Orbit Coupling,
                   \textit{ibid.} \textbf{115}, 106401 (2016);
                K. Wohlfeld, S.~Nishimoto, M.~W.~Haverkort,
                   and J. van den Brink,
                   Microscopic origin of spin-orbital separation in Sr$_2$CuO$_3$,
                   Phys. Rev. B \textbf{88}, 195138 (2013);
                C.-C. Chen, M. van Veenendaal, T.~P.~ Devereaux,
                   and K. Wohlfeld,
                   Fractionalization, entanglement, and separation:
                   Understanding the collective excitations
                   in a spin-orbital chain,
                   \textit{ibid.} \textbf{91}, 165102 (2015).

\bibitem{Brz15} W. Brzezicki, A. M. Ole\'s, and M. Cuoco,
                   Spin-Orbital Order Modified by Orbital Dilution
                   in Transition-Metal Oxides: From Spin Defects
                   to Frustrated Spins Polarizing Host Orbitals,
                   Phys. Rev. X \textbf{5}, 011037 (2015);
                W.~Brze\-zicki, M. Cuoco, and A. M. Ole\'s,
                   Novel spin-orbital phases induced by orbital dilution,
                   J. Supercond. Nov. Magn. \textbf{29}, 563 (2016).

\bibitem{Ole12} A. M. Ole\'s,
                   Fingerprints of spin-orbital entanglement
                   in transition metal oxides,
                   J. Phys.: Condens. Matter \textbf{24}, 313201 (2012);
                   Frustration and entanglement in compass and spin-orbital models,
                   Acta~Phys. Polon. A \textbf{127}, 163 (2015).

\bibitem{Rex12} Rex Lundgren, V. Chua, and G. A. Fiete,
                   Entanglement entropy and spectra of
                   the one-dimensional Kugel-Khomskii model,
                   Phys. Rev. B \textbf{86}, 224422 (2012).

\bibitem{You15} W.-L. You, A. M. Ole\'s, and P. Horsch,
                   Entanglement driven phase transitions in spin-orbital models,
                   New J. Phys. \textbf{17}, 083009 (2015);
                W.-L. You, P. Horsch, and A. M. Ole\'s,
                   Quantum entanglement in the one-dimensional spin-orbital
                   SU(2)$\otimes$XXZ model,
                   Phys. Rev. B \textbf{92}, 054423 (2015).

\bibitem{Miy02} S. Miyasaka, Y. Okimoto, and Y. Tokura,
                   Anisotropy of Mott-Hubbard gap transitions due to
                   spin and orbital ordering in LaVO$_3$ and YVO$_3$,
                   J. Phys. Soc. Jpn. \textbf{71}, 2086 (2002).

\bibitem{Fuj10} J. Fujioka, T. Yasue, S. Miyasaka, Y. Yamasaki, T.~Arima,
                   H. Sagayama, T. Inami, K. Ishii, and Y. Tokura,
                   Critical competition between two distinct
                   orbital-spin ordered states in perovskite vanadates,
                   Phys. Rev. B \textbf{82}, 144425 (2010).

\bibitem{Dag01} E. Dagotto, T. Hotta, and A. Moreo,
                   Colossal magnetoresistant materials:
                   The key role of phase separation,
                   Phys. Rep. \textbf{344}, 1 (2001);
                E. Dagotto,
                   Open questions in CMR manganites, relevance of
                   clustered states and analogies with other compounds
                   including the cuprates,
                   New J. Phys. \textbf{7}, 67 (2005).

\bibitem{Tok06} Y. Tokura,
                   Critical features of colossal magnetoresistive manganites,
                   Rep. Prog. Phys. \textbf{69}, 797 (2006).

\bibitem{Bal10} Leon Balents,
                   Spin liquids in frustrated magnets,
                   Nature (London) \textbf{464}, 199 (2010).

\bibitem{Bal16} Lucile Savary and Leon Balents,
                   Quantum spin liquids,
                   arXiv:1601.03742 (2016).

\bibitem{Fei05} L. F. Feiner and A. M. Ole\'s,
                   Orbital liquid in ferromagnetic manganites:
                   The orbital Hubbard model for eg electrons,
                   Phys. Rev. B \textbf{71}, 144422 (2005);
                A. M. Ole\'s and L. F. Feiner,
                   Why spin excitations in ferromagnetic manganites are isotropic,
                   \textit{ibid.} \textbf{65}, 052414 (2002).

\bibitem{Nor08} B. Normand and A. M. Ole\'s,
                   Frustration and entanglement in the $t_{2g}$
                   spin-orbital model on a triangular lattice:
                   Valence-bond and generalized liquid states,
                   Phys. Rev. B \textbf{78}, 094427 (2008);
                B. Normand,
                   Multicolored quantum dimer models, resonating valence-bond states,
                   color visons, and the triangular-lattice $t_{2g}$ spin-orbital system,
                   \textit{ibid.} \textbf{83},  064413 (2011);
                J. Chaloupka and A. M. Ole\'s,
                   Spin-orbital resonating valence bond liquid on a triangular
                   lattice: Evidence from finite-cluster diagonalization,
                   \textit{ibid.} \textbf{83}, 094406 (2011).

\bibitem{Karlo} P. Corboz, M. Lajk\'o, A. M. La\"uchli, K. Penc, and F.~Mila,
                   Spin-Orbital Quantum Liquid on the Honeycomb Lattice,
                   Phys. Rev. X \textbf{2}, 041013 (2012).

\bibitem{Nas12} J. Nasu and S. Ishihara,
                   Dynamical Jahn-Teller effect in a spin-orbital
                   coupled system,
                   Phys. Rev. B \textbf{88}, 094408 (2013).

\bibitem{Sel14} E. Sela, H.-C. Jiang, M. H. Gerlach, and S. Trebst,
                   Order-by-disorder and spin-orbital liquids
                   in a distorted Heisenberg-Kitaev model,
                   Phys. Rev. B \textbf{90}, 035113 (2014).

\bibitem{Mil14} A. Smerald and F. Mila,
                   Exploring the spin-orbital ground state of Ba$_3$CuSb$_2$O$_9$,
                   Phys. Rev. B \textbf{90}, 094422 (2014);
                   Disorder-Driven Spin-Orbital Liquid Behavior
                   in the Ba$_3$XSb$_2$O$_9$ Materials,
                   Phys. Rev. Lett. \textbf{115}, 147202 (2015).

\bibitem{Ver04} F. Vernay, K. Penc, P. Fazekas, and F. Mila,
                   Orbital degeneracy as a source of frustration in LiNiO$_2$,
                   Phys. Rev. B \textbf{70}, 014428 (2004).

\bibitem{Brz12} W. Brzezicki, J. Dziarmaga, and A. M. Ole\'s,
                   Noncollinear magnetic order stabilized by entangled
                   spin-orbital fluctuations,
                   Phys. Rev. Lett. \textbf{109}, 237201 (2012);
                   Exotic Spin Orders driven by orbital fluctuations
                   in the Kugel-Khomskii Model,
                   Phys. Rev. B \textbf{87}, 064407 (2013);
                   Exotic spin order due to orbital fluctuations,
                   Acta Phys. Polon. A \textbf{126}, A-40 (2014).

\bibitem{Rei05} A. Reitsma, L. F. Feiner, and A. M. Ole\'s,
                   Orbital and spin physics in LiNiO$_2$ and NaNiO$_2$,
                   New J. Phys. \textbf{7}, 121 (2005).

\bibitem{Goo55} J. B. Goodenough,
                   Theory of the Role of Covalence in the
                   Perovskite-Type Manganites [La,M(II)]MnO$_3$,
                   Phys. Rev. \textbf{100}, 564 (1955).

\bibitem{Fei98} D. Feinberg, P. Germain, M. Grilli, and G. Seibold,
                   Joint superexchange-Jahn-Teller mechanism
                   for layered antiferromagnetism in LaMnO$_3$,
                   Phys. Rev. B \textbf{57}, R5583 (1998).

\bibitem{Nan10} B. R. K. Nanda and S. Satpathy,
                   Magnetic and orbital order in LaMnO$_3$ under uniaxial strain:
                   A model study,
                   Phys. Rev. B \textbf{81}, 174423 (2010).

\bibitem{Kov10} N. N. Kovaleva, A. M. Ole\'s, A. M. Balbashov, A. Maljuk,
                   D. N. Argyriou, G. Khaliullin, and B. Keimer,
                   Low-energy Mott-Hubbard excitations in LaMnO$_3$
                   probed by optical ellipsometry,
                   Phys. Rev. B \textbf{81}, 235130 (2010).

\bibitem{Fei99} L. F. Feiner and A. M. Ole\'s,
                   Electronic origin of magnetic and orbital
                   ordering in insulating LaMnO$_3$,
                   Phys. Rev. B \textbf{59},   3295 (1999).

\bibitem{vdB99} J. van den Brink, P. Horsch, F. Mack, and A. M. Ole\'s,
                   Orbital dynamics in ferromagnetic transition metal oxides,
                   Phys. Rev. B \textbf{59}, 6795 (1999).

\bibitem{Pav10} Eva Pavarini and Erik Koch,
                   Origin of Jahn-Teller distortion and orbital order in LaMnO$_3$,
                   Phys. Rev. Lett. \textbf{104}, 086402 (2010).

\bibitem{Ole00} A. M. Ole\'s, L. F. Feiner, and J. Zaanen,
                   Quantum Melting of Magnetic Long-Range Order near Orbital
                   Degeneracy: I. Classical Phases and Gaussian Fluctuations,
                   Phys. Rev. B \textbf{61}, 6257 (2000).

\bibitem{Hal71} B. Halperin and R. Englman,
                   Cooperative dynamic Jahn-Teller effect.
                   II. Crystal distortions in perovskites,
                   Phys. Rev. B \textbf{3}, 1698 (1971).

\bibitem{Geh75} G. A. Gehring and K. A. Gehring,
                   Co-operative Jahn-Teller effects,
                   Rep. Prog. Phys. \textbf{38}, 1 (1975).

\bibitem{Oka02} S. Okamoto, S. Ishihara, and S. Maekawa,
                   Orbital ordering in LaMnO$_3$:
                   Electron-electron and electron-lattice interactions,
                   Phys. Rev. B  \textbf{65}, 144403 (2002).

\bibitem{Sik03} O. Sikora and A. M. Ole\'s,
                   Origin of the Orbital Ordering in LaMnO$_3$,
                   Acta Phys. Polon. B \textbf{34}, 861 (2003).

\bibitem{Huang97} Q. Huang, A. Santoro, J. W. Lynn, R. W. Erwin,
                   J.~A.~Borchers, J. L. Peng, and R. L. Greene,
                   Structure and magnetic order in undoped lanthanum
                   manganite,
                   Phys. Rev. B \textbf{55}, 14987 (1997).

\bibitem{Alb11} A. F. Albuquerque, D. Schwandt, B. Het\'enyi, S. Capponi,
                   M. Mambrini, and A. M. L\"auchli,
                   Phase diagram of a frustrated quantum antiferromagnet
                   on the honeycomb lattice: Magnetic order versus
                   valence-bond crystal formation,
                   Phys. Rev. B  \textbf{84}, 024406 (2011).

\bibitem{Got16} D. Gotfryd, J. Rusna\v{c}ko, K. Wohlfeld, G. Jackeli,
                   J.~Chaloupka, and A. M. Ole\'s,
                   Phase diagram and spin correlations of the Kitaev-Heisenberg
                   model: Importance of quantum effects,
                   arXiv:1608.05333 (2016).

\bibitem{Fle04} This formula is based on experimental data and
                   was introduced by G. S. Rushbrooke and P. J. Wood,
                   Mol. Phys. \textbf{1}, 257 (1958); see also:
                   M. Fleck, A. I. Lichtenstein, M. G. Zacher, W. Hanke,
                   and A. M. Ole\'s,
                   On the nature of the magnetic transition in a Mott
                   insulator, Eur. Phys. J. B \textbf{37}, 439 (2004).

\bibitem{Ryn10} A. van Rynbach, S. Todo, and S. Trebst,
                   Orbital Ordering in $e_g$ Orbital Systems:
                   Ground States and Thermodynamics of the 120\textdegree\ Model,
                   Phys. Rev. Lett. \textbf{105}, 146402 (2010).

\bibitem{Cza16} P. Czarnik, J. Dziarmaga, and A. M. Ole\'s,
                   Variational tensor network renormalization in imaginary time:
                   Two-dimensional quantum compass model at finite temperature,
                   Phys. Rev. B \textbf{93}, 184410 (2016).

\bibitem{Ole06} A. M. Ole\'s, P. Horsch, L. F. Feiner, and G. Khaliullin,
                   Spin-Orbital Entanglement and Violation
                   of the Goodenough-Kanamori Rules,
                   Phys. Rev. Lett. \textbf{96}, 147205 (2006).

\bibitem{Kru09} F. Kr\"uger, S. Kumar, J. Zaanen, and J. van den Brink,
                   Spin-orbital frustrations and anomalous metallic state
                   in iron-pnictide superconductors,
                   Phys. Rev. B \textbf{79}, 054504 (2009).

\bibitem{DiP15} P. Di Pietro, J. Hoffman, A. Bhattacharya, S. Lupi, and A. Perucchi,
                   Spectral Weight Redistribution in (LaNiO$_3)_n$/(LaMnO$_3)_2$
                   Superlattices from Optical Spectroscopy,
                   Phys. Rev. Lett. \textbf{114}, 156801 (2015).






\end{thebibliography}
\end{document}